\documentclass[twoside,11pt]{article}

\usepackage{jmlr2e}
\usepackage{amsmath}
\usepackage{xcolor}
\usepackage[utf8]{inputenc}
\usepackage{lineno}

\DeclareMathOperator*{\argmin}{arg\,min}

\newcommand{\bsigma}{{\boldsymbol{\sigma}}}
\newcommand{\T}{{\mathcal{T}}}
\newcommand{\tT}{{t\in\T}}

\newcommand{\I}{{\mathcal{I}}}
\newcommand{\iI}{{i\in\I}}

\newcommand{\B}{{\mathcal{B}}}
\newcommand{\bB}{{b\in\B}}

\newcommand{\M}{{\mathcal{M}}}
\newcommand{\iM}{{i\in\M}}

\newcommand{\K}{{\mathcal{K}}}
\renewcommand{\S}{{\mathcal{S}}}
\newcommand{\X}{{\mathcal{X}}}

\newcommand{\F}{{\mathcal{F}}}
\newcommand{\N}{{\mathcal{N}}}
\newcommand{\R}{{\mathds{R}}}

\newcommand{\btsigma}{{\widetilde{\boldsymbol\sigma}}}
\newcommand{\tsigma}{{\widetilde{\sigma}}}
\newcommand{\ttau}{{\widetilde{\tau}}}
\newcommand{\tdelta}{{\widetilde{\Delta}}}

\renewcommand{\P}{{\mathbb{P}}}

\usepackage{todonotes}

\usepackage{array}
\usepackage{booktabs}
\usepackage{dsfont}
\usepackage{algorithm}
\usepackage{algpseudocode}
\algnewcommand\algorithmicinput{\textbf{Input:}}
\algnewcommand\Input{\item[\algorithmicinput]}
\algnewcommand\algorithmicoutput{\textbf{Output:}}
\algnewcommand\Output{\item[\algorithmicoutput]}

\title{FuNVol: A Multi-Asset Implied Volatility Market Simulator using Functional Principal Components and Neural SDEs
\thanks{A github repository that generates dynamical IV surfaces may be found at \url{https://github.com/vedantch/FuNVol}.}}

\begin{document}

\author{\name Vedant Choudhary \email vedant.choudhary@mail.utoronto.ca \\
       \addr Department of Statistical Sciences\\
       University of Toronto\\
       Toronto, ON M5G 1Z5, Canada
       \AND
       \name Sebastian Jaimungal\email sebastian.jaimungal@utoronto.ca \\
        \addr Department of Statistical Sciences\\
       University of Toronto\\
       Toronto, ON M5G 1Z5, Canada
       \AND
       \name Maxime Bergeron\email mb@riskfuel.com \\
        \addr Riskfuel Analytics\\
       Toronto, ON, Canada}

\editor{}

\maketitle

\begin{abstract}
We introduce a new approach for generating sequences of implied volatility (IV) surfaces across multiple assets that is faithful to historical prices. We do so using a combination of functional data analysis and neural stochastic differential equations (SDEs) combined with a probability integral transform penalty to reduce model misspecification. We demonstrate that learning the joint dynamics of IV surfaces and prices produces  market scenarios that are consistent with historical features and lie within the sub-manifold of surfaces that are essentially free of static arbitrage. Finally, we demonstrate that delta hedging using the simulated surfaces generates profit and loss (P\&L) distributions that are consistent with realised P\&Ls.
\end{abstract}


\section{Introduction}

It is well known that implied volatility (IV) surfaces exhibit skews/smiles and term structures that are inconsistent with the na\"ive assumptions of Black-Scholes. Moreover, IV surfaces evolve over time as the underlying asset price fluctuates and traders incorporate  market information into option prices. These surface fluctuations pose challenges to modeling. Furthermore, IV surfaces must be free of static arbitrage, which imposes constraints on the shape of the IV surfaces \citep{gatheral2014arbitrage, durrleman2010implied, roper2010arbitrage}, and adds to the complexity of modeling them.

Leaving aside the modeling of the evolution dynamics, even the problem of interpolating arbitrage-free IV surfaces is not straightforward. Markets have option price quotes for only a finite (and often limited) number of times to expiry and strikes, and extending and interpolating the surfaces to a continuum is nontrivial. In the extant literature, there are a variety of approaches to fit continuous IV surfaces to observed discrete data, including spline interpolation \citep{fengler2009arbitrage, orosi2012empirical}, parametric models such as surface SVI \citep{gatheral2014arbitrage} and SABR \citep{hagan2002managing}, and non-parametric approaches \citep{cont2002dynamics, fengler2003fitting, fengler2003dynamics}. These approaches, while useful, have some limitations. For instance, they may impose overly strong assumptions on the shape of IV surfaces, consequently, good fits to historical data may be difficult and the resulting fits may introduce arbitrage. To avoid some of these limitations, researchers are adopting machine learning techniques  to model IV surfaces, for example: support vector machines \citep{zeng2019online} and variational autoencoders \citep{ning2021arbitrage, bergeron2022variational}. To prevent static arbitrage in the IV surfaces, \cite{ackerer2020deep} penalize the arbitrage constraints whereas \cite{zheng2021incorporating} impose hard constraints on the neural network architecture. The advantage of these deep learning approaches is that they replicate market data better, especially when data is sparse. Additionally, they are able to embed financial constraints on IV surfaces through appropriate activation functions, neural network architectures, and loss functions. 

Shifting from a static perspective to a dynamic one, \cite{cont2002dynamics} study IV surfaces through the lens of the Karhunen-Lo\'eve decomposition of their daily variations. \cite{cont2002stochastic} build on this to jointly model the evolution of the IV surface and the underlying. \cite{bloch2021deep} model IV surface dynamics by introducing  stochastic evolution of the parameters of the SVI model \citep{gatheral2004parsimonious}. They use deep learning (with architecture comprising convolutional LSTMs \citep{shi2015convolutional}) to capture the spatiotemporal relationship between the strikes and expiries. \cite{shang2022dynamic} employ dynamic functional principal component analysis (FPCA) to forecast IV smiles (across deltas for a fixed maturity) in foreign exchange (FX) markets, using ARIMA to forecast the functional principal component coefficients (FPCCs). These works focus on one-step ahead point forecasts of IV smiles or surfaces, and are not full generative models. 

In contrast with these works, we develop a conditional generative model that captures the historical temporal dynamics of the IV surfaces  and allows us to simulate sequences of IV surfaces over multiple steps into the future. Furthermore, our proposed modeling paradigm allows us to incorporate (and simulate) other market observables such as asset prices, interest rates, trade volume, and VIX. The framework is very flexible and has many applications. For example, the generative model may be used to simulate a large number of sequences of IV surfaces, and any associated market data, which can then be fed as state inputs to train data-hungry reinforcement learning (RL) algorithms for hedging \citep{buehler2019deep, cao2021deep}, portfolio allocation \citep{coache2021reinforcement, coache2022conditionally, wang2022objective}, statistical arbitrage strategies \cite{zhao2022deep}, and so on. We demonstrate the performance of our model by generating sequences of IV surfaces and corresponding asset prices for four popular equity tickers. 

Our model does not assume a parametric form for surfaces such as SSVI or SABR. Instead, we use functional data analysis \citep{ramsaysilverman} in the spirit of \cite{cont2002dynamics} (who, however, model changes in log-IV surfaces) to provide a direct functional representation of IV surfaces and then model their evolution with non-Markovian neural SDEs \citep{kidger2020neural, chen2018neural, tzen2019neural, gierjatowicz2020robust, cuchiero2020generative}. Given the IVs on a discrete grid of option deltas and maturities, we project the data onto a set of Legendre basis functions to generate a complete surface. The Legendre basis functions provide good surface fits and have the added advantage of being orthogonal which simplifies the task of Functional Principal Component Analysis (FPCA). Our modeling approach, however, is agnostic to the choice of basis functions. As a result, the main sources of variation in the surfaces (across all assets) are represented by the Functional Principal Components (FPCs). Each IV surface is then projected onto a lower dimensional set of FPCs, providing us with a time series of Functional Principal Component Coefficients (FPCCs) corresponding to IV surface evolution. In particular, we obtain a low dimensional model-agnostic characterization of complete IV surfaces. 

\begin{figure} [h!]
    \centering
    \includegraphics[width=0.95\columnwidth]{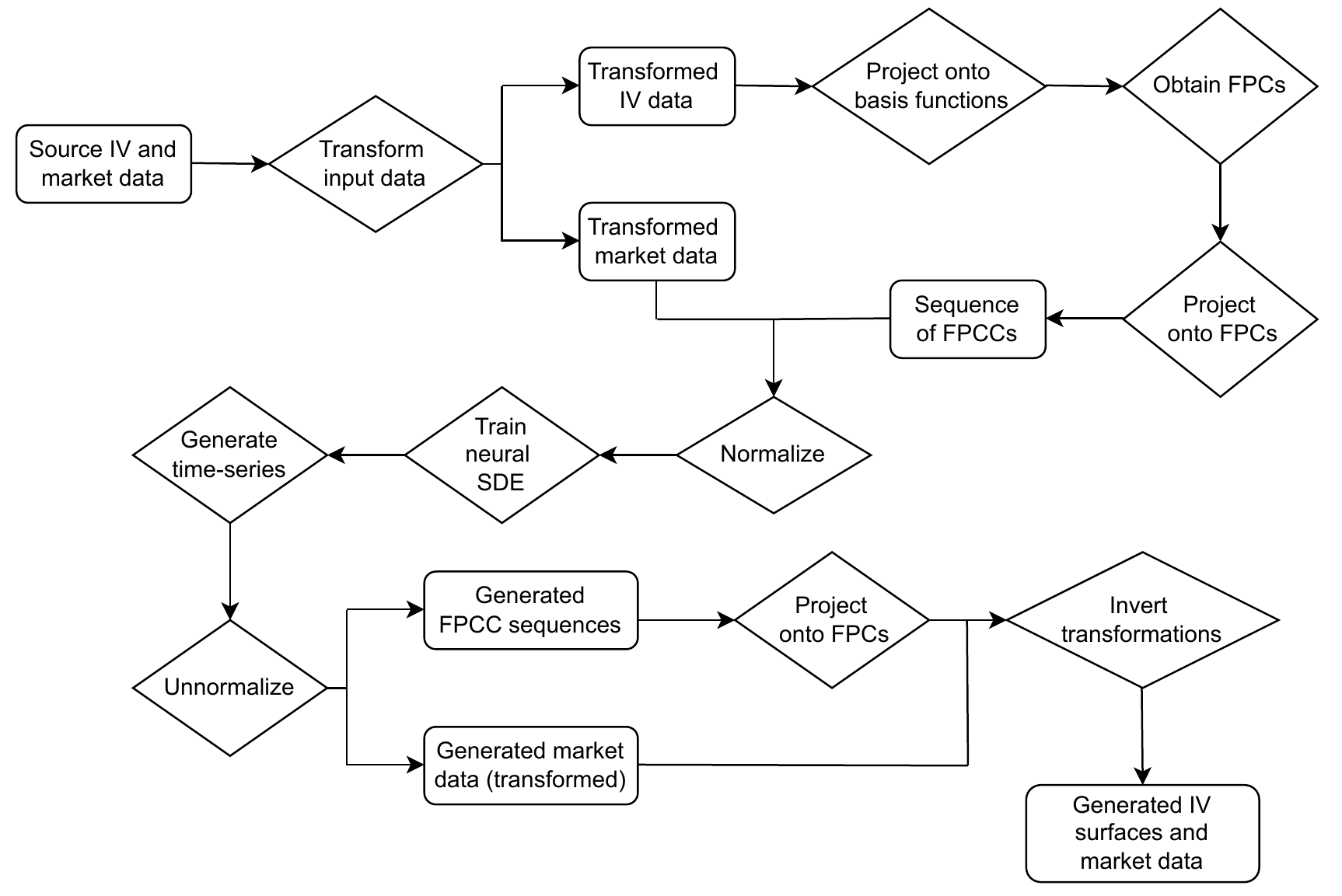}
    \caption{Flowchart illustrating the end-to-end process to generate synthetic IV surfaces and market data.} 
    \label{fig:flowchart}
\end{figure}

The temporal dynamics of the FPCCs are captured by a neural SDE whose drift and diffusion are parametrized by neural networks comprising Gated Recurrent Units (GRUs) \citep{cho2014learning} coupled to feed forward (FF) layers. The choice of GRUs was driven by their ability to learn long range dependencies, with potentially fewer parameters than long short term memory (LSTM) layers. The time series of other relevant market data can also be incorporated into the neural SDE at this stage to simulate consistent market scenarios. Modeling the time series using a data-driven approach, like neural SDEs, rather than a parametric model like ARIMA or GARCH has the added advantage of reducing model misspecification. We impose a novel probability integral transform (PIT) penalty to reduce model misspecification even further, which to the best of our knowledge is the first of its kind. Static arbitrage constraints can be embedded into the neural SDE loss function, however, we demonstrate that the generated surfaces are mostly arbitrage-free. Specifically, we show using static arbitrage metrics that the generated surfaces have less overall static arbitrage than the market data which is used for training. Hence, we  do not explicitly impose penalties for static arbitrage. A flowchart of the entire process may be viewed as in Figure \ref{fig:flowchart}. Our main contribution is the development of an approach that consistently models the evolution of multiple assets' IV surfaces, together with other relevant market data, in a model-agnostic manner.

The paradigm of generative modeling using deep learning has gained immense traction for the past decade since the seminal work of \cite{goodfellow2020generative} on Generative Adversarial Networks (GANs). The aim of generative modeling is to implicitly learn the distribution of the observed data by generating samples from the target distribution. Training these models requires huge amounts of data, which while easily available for visual or audio data, is not readily available when it comes to time series data. This data scarcity is especially true in a financial context where just one realization of a stochastic process is observed, and the non-stationarity of market data means only a small subset of it is relevant for the modeling exercise in the present. The lack of consensus regarding reliable test metrics to assess the quality of generated time series data presents another challenge. 

Generative models that specifically focus on IV surfaces have garnered considerable attention lately. \cite{carmona2017simulation} propose an arbitrage-free Monte-Carlo simulator for future paths of IV surfaces that is consistent with historical data based on tangent L\'evy models. \cite{wiese2019deep} use GANs to simulate prices of equity index options on a discrete grid of strike and maturity by working with the discrete local volatilities instead of IVs to avoid complex arbitrage constraints. \cite{cuchiero2020generative} develop a GAN-based model for calibrating, on a fixed calendar date, a local stochastic volatility model parametrised by neural networks. \cite{cohen2021arbitrage} develop a factor-based model for European options in which they extract latent market factors from finitely many option prices directly rather than from IVs as we do. The factor dynamics are modeled via neural SDEs, with the drift and diffusion neural networks having hard constraints to prevent static arbitrage. \cite{cont2022simulation} propose VolGAN that learns the one-step ahead joint conditional distribution of the changes in the log IV surfaces, the underlying return conditional on the realized volatility, the IV surface of the previous day, and the price returns for the past two days. It, however, generates the IVs on a fixed moneyness-maturity grid rather than modeling the entire surface. They also propose a second approach based on FPCA of the changes of the log IV surfaces and model the coefficients by Ornstein-Uhlenbeck processes, identical to what was done in \cite{cont2002dynamics}, but now including a down-sampling method for reducing static arbitrage. \cite{francois2023joint} model the joint dynamics of the IV surfaces and its underlying using a five-factor parametric representation of the IV surfaces. The underlying returns and the factor coefficients are modeled using GARCH-type models, with a Gaussian copula capturing the dependence structure. Our work contrasts with the extant literature, as we take a model-agnostic approach and use functional data analysis and neural-SDEs to characterize the dynamics of the entire surface together with other market factors.  

The remainder of the paper is structured as follows. In Section \ref{sec:fda}, we discuss the projection of market data on a set of Legendre basis functions followed by a FPCA of the data. The training of a neural SDE model on the FPCCs is presented in Section \ref{sec:neural_sde}. Section \ref{sec:quant-arb} deals with the metrics to quantify the amount of arbitrage in the generated surfaces in contrast to the observed ones. Finally, the results, together with a delta hedging exercise, are presented in Section \ref{sec:results}.

\section{\textbf{Functional Data Projection}}\label{sec:fda}

Vanilla financial options are liquid for only a limited collection of strikes (or deltas) and maturities. The pricing of exotic options and dynamic hedging, however requires the entire volatility surface. In the extant literature, there exists several methods for obtaining an IV surface from discrete values, including parametric and non-parametric techniques. Parametric approaches attempt to fit the discrete data points using a parametric model, which can then be used to interpolate and extrapolate to the whole IV surface. Common non-parametric approaches include using Nadaraya-Watson kernel estimators and FPCA \citep{shang2022dynamic, benko2009common}. Here, we model the whole IV surface using two-dimensional normalized Legendre polynomials as basis functions followed by FPCA to reduce the dimensionality. The basis functions in our case are product of Legendre polynomials parameterized by delta and maturity which enables us to obtain entire surfaces in these parameters rather than just IV smiles requiring further interpolation. Using FPCs as factors to model the IV surfaces enables us to capture the most important co-movements of IV across maturities and deltas simultaneously.

\subsection{Basis Function Projection}\label{sec:meth_basis_proj}

We denote the sequence of historical IV surfaces  by $\{\bsigma_t\}_{\tT}$, where $\T:=\{1,\dots,T\}$ denotes the collection of historical dates on which we have data. On each date $t\in\T$, a given IV surface is represented by $\bsigma_t:=(\sigma_{t}(\tau_i, \Delta_i))_{\iI_t}$ where $\I_t:=\{i,\dots,I_t\}$ -- we may often use the more compact notation  $\sigma_{t,i}:=\sigma_t(\tau_i,\Delta_i)$. Thus, $(\tau_i,\Delta_i)_{\iI_t}$ represents the collection of pairs of time to maturity and delta that IV data is available on day $t$. In principle, the $(\tau,\Delta)$ pairs do not need to lie on a grid, nor do they need to be the same collection of pairs across all days. In practice, however, it may be convenient to have a fixed grid through time. 

Our first goal is to obtain a faithful representation of the IV surfaces across all days using a functional basis in $(\tau,\Delta)$ space. The orthogonality of the basis is a desirable property as it simplifies the task of FPCA which we perform later. This is one of the advantages of choosing Legendre polynomials as basis over, e.g., B-splines. In our implementations, we use normalized Legendre polynomials to generate the basis. Recall that one representation of (normalized) Legendre polynomials of order-$m$ is $L_m(x):=\frac{\sqrt{2m+1}}{2}\,\frac{1}{2^mm!}\frac{d^m}{dx^m}(x^2-1)^m$. To this end, we denote the set of basis functions by $(\phi_b)_\bB$ with $\B:=\{1,\dots,B\}$, where $\phi_b:\X\to\R$ and $\X:=[-1,1]\times[-1,1]$.  Specifically, our set of basis functions $(\phi_b)_\bB$ is given by $\{L_i(\tau) L_j(\Delta): 0 \leq i+j \leq n_o;\ i,j\geq0 \}$ where $\B$ is an enumeration of the set $\{0 \leq i+j \leq n_o;\ i,j\geq0\}$. We find that $n_o=4$ provides good surface fits to the data and hence obtain $B=\frac12(n_o+1)(n_o+2)=15$ basis functions. In other words, we take all products of the first $n_o$ order Legendre polynomials in $\tau$ and $\Delta$ such that the powers sum to at most $n_o$ and where one Legendre polynomial applies to $\tau$ and the other to $\Delta$.

While the $(\tau,\Delta)$ pairs from data do not lie in $\X$ and $\bsigma$ are all positive, we first pre-process the data, as described in detail in \ref{sec:res_data_trans}, to map $(\tau,\Delta)$ into $\X$ and $\bsigma$ into $\R$. We denote the transformed data as $\btsigma$, $\ttau$, and $\tdelta$. With the hope that no confusion shall arise, we omit the tilde notation in this subsection from this point so that $\sigma_t, \tau$ and $\Delta$ refer to $\tsigma_t, \ttau$ and $\tdelta$ respectively. 

The normalized Legendre polynomials  are orthonormal in the interval $[-1,1]$ in the following sense:
\begin{equation}
    \int_{-1}^1 L_n(x) L_m(x) dx =\delta_{nm}\,,\qquad \forall n,m
\end{equation}
where $\delta_{nm}$ is the Kronecker delta -- which equals zero unless $n=m$, in which case it equals one. The orthonormality of the Legendre polynomials on $[-1,1]$ is inherited by our choice of basis functions on the space $\X$. In particular we have that
\begin{equation}
    \int_{-1}^1 \int_{-1}^1 \phi_n(x,y) \phi_m(x,y)\ dx\ dy = \delta_{nm}\,,
    \qquad\forall n,m\in\B\,.
\end{equation}

Given IV data, $(\bsigma_t)_\tT$, for each day $t\in\T$, we project its transformed version $(\btsigma_t)_\tT$ onto the chosen basis to generate a continuous surface $\check{\sigma}_t:\X\to\R$ as follows
\begin{equation}
    \check{\sigma}_t(x,y) = \sum_{k\in\B} a_{t,k}\,\phi_k(x,y)\,, \qquad (x,y)\in\X
\end{equation}
where the sequence of coefficients $((a_{t,k})_{k\in\B})_\tT$ are estimated by minimizing the least-square errors 
\begin{equation}\label{eqn:sse}
\boldsymbol{a}_{t}=\argmin_{\boldsymbol{\alpha}_{t}}  \sum_{\iI_t} \left(\sigma_{t,i} - \sum_{k\in\B} \alpha_{t,k}\ \phi_k(\tau_i, \Delta_i)\right)^{2}\,.
\end{equation}
We assume that $|\I_t|>|\B|$ so that the regression is well posed. When working with multiple assets, the above optimization is carried out separately for each asset (albeit using the same basis functions), which leads to a distinct sequence of estimated coefficients  for each asset $((a_{t,k})_{k\in\B})_\tT$. 

\begin{figure} [h!]
    \centering
    \includegraphics[width=0.95\columnwidth]{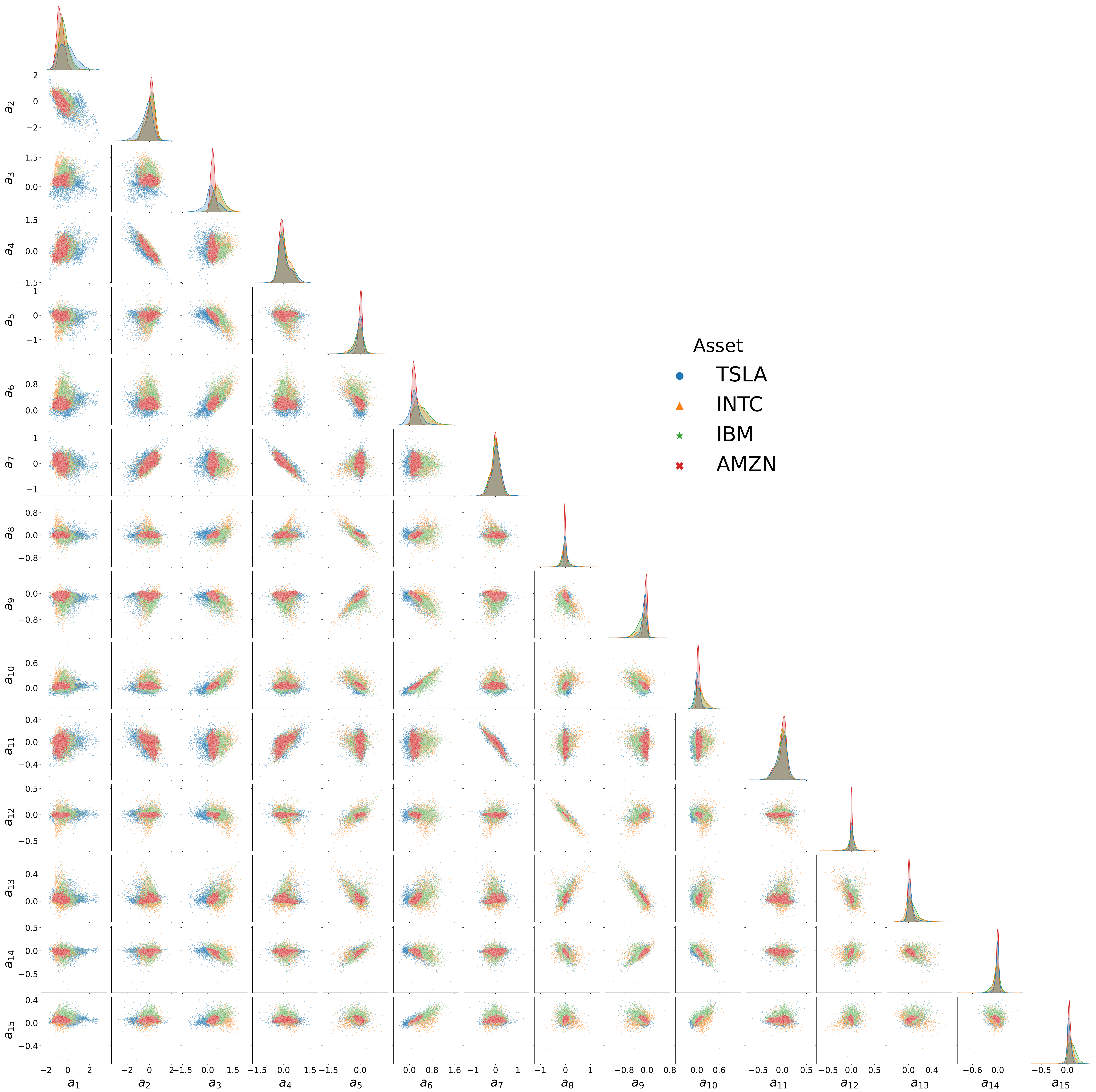}
    \caption{Pairwise scatter plot of the estimated time series of coefficients $\{a_{t,k}\}$ corresponding to the projection of IV data on the orthonormalized Legendre basis functions.} 
    \label{fig:pairwise_scatter_a}
\end{figure}

Figure \ref{fig:pairwise_scatter_a} shows a scatter plot of all pairs of the estimated coefficients for the four sets of IV surfaces that we analyse. The figure illustrates that there is a high degree of dependence between the coefficients, and this motivates us to perform a dimensional reduction on the functional regression by making use of FPCA. For a review of FPCA we refer to  \cite{ramsaysilverman}. 
In the next subsections we provide some of the key results needed to develop the FPCs.

\subsection{Obtaining functional principal components}\label{sec:meth_obtain_fpc}

We next summarize some of the key results needed to construct FPCs.
Consider a compact set $\S \subset \R^m$ and let $K: \S \times \S \rightarrow \R$ be a symmetric function. Define the corresponding kernel operator acting on square-integrable functions $f \in L^2(\S)$ as $\K: L^2(\S) \rightarrow L^2(\S)$ given by
\begin{equation}\label{eqn:eigen_eqn}
    (\mathcal{K} f)(u) = \int_\S K(u,v)\ f(v)\ dv
\end{equation}
The kernel operator is said to be positive semi-definite if $\forall f \in L^2(\S)$, we have 
\begin{equation}
    \int_{\S\times \S} K(u,v)\ f(u)\ f(v)\ du\ dv \geq 0
\end{equation}

\begin{theorem}[Mercer's theorem \citep{steinwart2012mercer}]\label{thm:mercer}
For symmetric functions $K(u,v)$ that are continuous on $\S \times \S$ and corresponding kernel operators $\K$ that are positive semi-definite, there exists an orthonormal sequence of continuous functions $\{\psi_i \in L^2(\S), i=1,2,\ldots\}$ which are eigenfunctions of $\K$ and eigenvalues $\{\lambda_i\geq 0, i=1,2,\ldots\}$ such that $\forall x \in \S$, $(\mathcal{K} \psi_i)(u) = \lambda_i \psi_i(u)$. Moreover, the function $K$ can be written in terms of the eigenpair as
\begin{equation}
    K(u,v) = \sum_{i=1}^{\infty} \lambda_i\ \psi_i(u)\ \psi_i(v)
\end{equation}
\end{theorem}
\noindent
As the eigenfunctions are orthonormal, we further have
\begin{equation}
    \int_\S \psi_i(u)\ \psi_j(u)\ du = \delta_{ij}\,,
\end{equation}
and hence
\begin{equation}
    \sum_{i=1}^{\infty} \lambda_i = \int_\S K(u,u)\ du < \infty\,.
\end{equation}

\begin{theorem}[Karhunen-Lo\'eve expansion \citep{karhunen1947under}]\label{thm:kar}
Consider a stochastic process $X: \S \times \Omega \rightarrow \R$ that is mean-square continuous with $X \in L^2(\S \times \Omega)$. Then, the eigenfunctions $\psi_i$ of the kernel operator $\K$ associated with the covariance function $K$ of the process $X$ act as a basis such that $\forall u \in \S$, the process can be represented as
\begin{equation}
    X(u) - \mathbb{E}[X(u)] = \sum_{i=1}^{\infty} \xi_i\ \psi_i(u) \in L^2(\Omega)\,,
\end{equation}
where the coefficients $\xi_i$ are random variables given by 
\begin{equation}
    \xi_i = \int_\mathcal{X} X(u)\ \psi_i(u)\ du\,.
\end{equation}
satisfying (i) $\mathbb{E}[\xi_i] = 0$ and (ii) $\mathbb{E}[\xi_i \xi_j] = \delta_{ij} \lambda_i$ where $\lambda_i$ is the eigenvalue corresponding to $\psi_i$. The infinite series representation of $X(u)$ converges uniformly on $\mathcal{X}$ with respect to the $L^2$ norm.
\end{theorem}
\noindent
In practice, the centered process $X$ is approximated using a dimensionally reduced basis consisting of $M$ eigenfunctions corresponding to the $M$ largest eigenvalues, i.e.,

\begin{equation}
    \widehat{X}(u) = \sum_{i=1}^{M} \xi_i\ \widehat{\psi}_i(u)\,
\end{equation}
and $\widehat{\psi}$ denotes the estimated eigenfunctions.

\begin{lemma}\label{lemma:merkar}
The covariance function of process $X$ given by $K(u,v)= \operatorname{Cov}(X(u),X(v))$ is continuous on $\S$ if and only if the process $X$ is mean-square continuous, i.e.
\begin{equation}
    \underset{\epsilon \rightarrow 0}{\lim}\ \mathbb{E}[(X(u+\epsilon)-X(u))^2] = 0\,.
\end{equation}
\end{lemma}
\noindent
The above lemma allows us to use the Karhunen-Lo\'eve expansion under the assumptions in Mercer's theorem. 
We next show how one estimates eigenfunctions and eigenvalues for a process $X$ given solely observations of process $\{X_1,\ldots,X_T\}$ when $\Omega = \T$. Further details on how we link this process to implied volatility surfaces in our case may be found in Section \ref{sec:meth_fpc_proj}. For this, we assume the data so that both the process and its observations are mean-centered.  If they are not, we apply  a shifting by their corresponding means.  First,  we  project the centered observations onto the chosen basis functions and express  the estimated eigenfunctions $\widehat{\psi}$ as a linear combination of the basis functions, as follows
\begin{equation}\label{eqn:basis}
    X_t(u) = \sum_{k=1}^B a_{t,k}\ \phi_k(u)   \quad \text{and} \quad  \widehat{\psi}_m(u) = \sum_{k=1}^B c_{m,k}\ \phi_k(u)\,.
\end{equation}
Denoting the matrix of time series of the coefficients by $A$, so that $A_{tk}=a_{t,k}$, the vector of basis functions by $\Phi = (\phi_1,\ldots,\phi_B)$, and the vector of the basis coefficients for the $m^{th}$ eigenfunction by $c_m=(c_{m,1},\ldots,c_{m,B})$, \eqref{eqn:basis} can be written in matrix notation as $X(u) = A\, \Phi(u)$ and $\widehat{\psi}_m(u) = \Phi(u)^\intercal c_m$. The empirical estimate of the covariance function $K$, denotes $\widehat{K}$, is given by the sample covariance of the observations
\begin{equation}
    \widehat{K}(u,v) = \tfrac{1}{T} \Phi(u)^\intercal A^\intercal A \Phi(v)\,.
\end{equation}
Using \eqref{eqn:eigen_eqn}, we get 
\begin{equation}
    (\mathcal{K} \widehat{\psi}_m)(u) = \int_\S \left[\tfrac{1}{T} \Phi(u)^\intercal A^\intercal A \Phi(v) \right] \Phi(v)^\intercal c_m\ dv = \tfrac{1}{T} \Phi(u)^\intercal A^\intercal A W c_m\,,
\end{equation}
where $W = \int_\S \Phi(v) \Phi(v)^\intercal\ dv$ is the basis weight matrix. Hence, we obtain the eigenproblem
\begin{equation}\label{eqn:eigen_problem}
    \tfrac{1}{T} \Phi(u)^\intercal A^\intercal A W c_m = \lambda_m \Phi(u)^\intercal c_m\,, \qquad \forall u \in \mathcal{X}\,.
\end{equation}
The eigenproblem can be solved by pre-multiplying \eqref{eqn:eigen_problem} by $W^{1/2}$ and integrating over the set $\S$ to get the finite dimensional problem
\begin{equation}\label{eqn:eigen_problem2}
    \left(\tfrac{1}{T} W^{1/2} A^\intercal A W^{1/2}\right) d_m = \lambda_m d_m\,,
\end{equation}
where $c_m = W^{-\frac12} d_m$. Working with orthonormal basis functions like normalized Legendre polynomials has the benefit that the basis weight matrix $W$ is identity. The eigenproblem in such a scenario reduces to PCA on the matrix of coefficients $A$. Once we obtain the solution to the eigenproblem in \eqref{eqn:eigen_problem2}, we can express the eigenfunctions in terms of the basis functions.

\subsection{Projecting onto functional principal components}\label{sec:meth_fpc_proj}

In our  application setting, $X=\sigma$, $\S = \X= [-1,1] \times [-1,1]$ corresponding to the grid range of $(\ttau, \tdelta)$ and $\Omega = \T$. During the functional principal component analysis, we treat our implied volatility as a random field indexed by the tuple $(\tau, \Delta) \in \mathcal{S}$, and each observation of the IV surface on a given day $t$ is treated as a realization of this random surface. In this section, when we speak of the IV surfaces we mean the transformed surfaces. With the hope that no confusion shall arise, we omit the tilde notation so that $\sigma_t$ here means $\tsigma_t$. Similarly, we refer to $(\ttau, \tdelta)$ simply by $(\tau, \Delta)$.
We choose the kernel function $K$ to be the covariance function of IV, i.e. $K(u,v) = \operatorname{Cov}(\bsigma(u), \bsigma(v))$ for $(u,v)\in\X\times\X$, which may be estimated by the sample covariance as
\begin{equation}
    \widehat{K}(u,v) = \tfrac{1}{T} \sum_{t=1}^T \left(\sigma_t(u) - \overline{\sigma}(u)\right) \left(\sigma_t(v) - \overline{\sigma}(v)\right)\qquad \text{ where } \qquad \overline{\sigma}(u) = \tfrac{1}{T} \sum_{t=1}^T \sigma_t(u)\,.
\end{equation}
The above formulation of the sample covariance and sample mean is for a single asset. It can be easily extended to the multi-asset case by having an additional summation running over the number of assets.

The sample covariance function is continuous with the corresponding kernel being positive semi-definite thus ensuring the applicability of Mercer's theorem. Note that here we treat the IV as a stochastic process indexed by maturity and delta, rather than time. As the assumptions of Mercer's theorem are satisfied by $\widehat{K}(u,v)$,  $\sigma(u)$ is mean-square continuous and hence the Karhunen-Lo\'eve expansion applies. 

\begin{figure} [h!]
    \centering
    \includegraphics[width=0.95\textwidth]{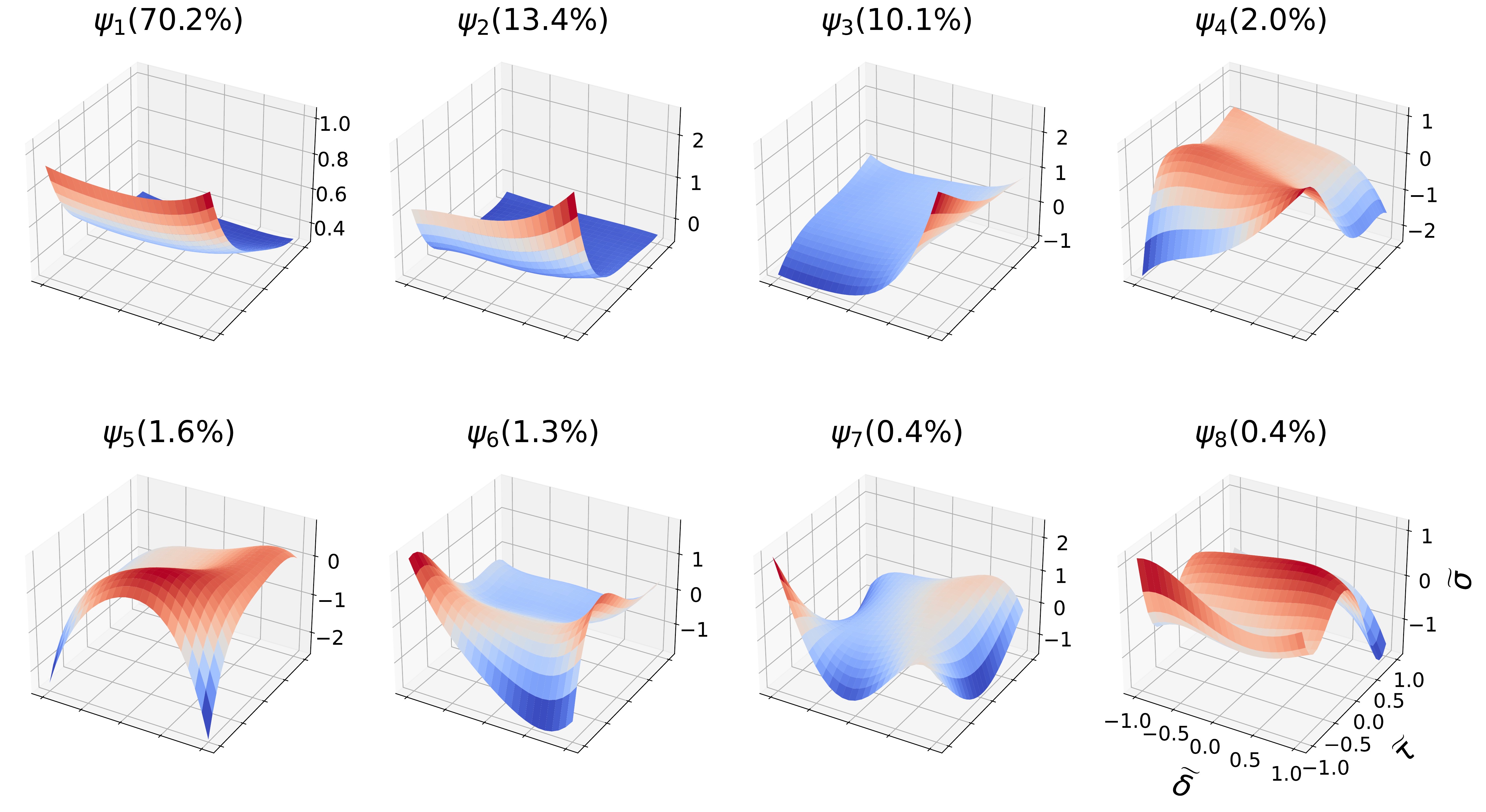}
    \caption{The eight largest functional principal components that explain at least $99.5\%$ of the variance in the IVs. The proportion of explained variance by each of them is in brackets. Note that the axes correspond to the transformed quantities  time to maturity, delta, and IV as detailed in Section \ref{sec:res_data_trans}.} 
    \label{fig:fpcs}
\end{figure}

We thus employ the Karhunen-Lo\'eve expansion to express the IV process in terms of the $B$ eigenfunctions obtained as in Section \ref{sec:meth_obtain_fpc}, and explicitly by solving the eigenproblem in \eqref{eqn:eigen_problem2}.  We do this using IVs from all assets after some transformations described in detail in Section \ref{sec:res_data_trans}. The resulting first $M=8$ (eight) eigenfunctions are shown in Figure \ref{fig:fpcs}. These eigenfunctions explain approximately $99.5\%$ of the variability of the surfaces, and their shape aligns well with the expected characteristics of the IV surfaces, namely term structure, skew, and convexity. Further details on the interpretation of the FPC surfaces can be found in Section \ref{sec:res_data_proj}.

To reduce the dimension of the data, we next project the data onto this reduced set of $M$ eigenfunctions and express the surfaces as
\begin{equation}
    \widehat\sigma_t(\tau, \Delta) = \sum_{m=1}^{M} b_{t,m}\ \psi_m(\tau, \Delta)\,.
    \label{eqn:sigma-from-psi}
\end{equation}
The sequence of coefficients $((b_{t,m}))$ are obtained by minimizing the sum of squared errors as in \eqref{eqn:sse}, but with the original basis functions replaced with the first $M$ eigenfunctions. Once again, the multi-asset case will have a distinct sequence of coefficients corresponding to each asset. We do not add any additional penalization, such as penalties associated with butterfly spread or calendar spread arbitrage, however, as we show in Section \ref{sec:generated-data}, our generator produces surfaces that are as free of arbitrage as the original surfaces.

Using this procedure, we obtain FPCCs as a time series $(b_t)_\tT$, $b_t\in \R^M$. In the next section, we proceed to model the time series of these FPCCs using neural SDEs. In this manner, we are able to construct dynamical surfaces from a low-dimensional set of coefficients, but in a manner that encodes path dependency.

\section{\textbf{Neural stochastic differential equations}} \label{sec:neural_sde}

In this section, we describe how we model the time series of FPCCs. Our key assumption on the dynamics is that FPCCs are driven by a neural SDE where the drift and diffusion terms do not necessarily have a Markovian structure, i.e., they may depend on the entire history of the coefficients, and furthermore, their values are obtained as outputs of neural networks. Consider an $M$-dimensional adapted Brownian motion  $W=(W_t)_{t\ge0}$ defined on the probability space $(\Omega, \F, \{\F_t\}_{t \ge0}, \P)$, with $\{\F_t\}_{t \ge0}$ the natural filtration generated by the Brownian motion. We assume that the FPCCs $(b_t)_{t\ge0}$ satisfy the SDE

\begin{equation}\label{eqn:SDE}
    db_t = \nu_t\ dt + \eta_t\ dW_t\,,
\end{equation}
where $\nu_t \in \R^M$ and $\eta_t \in \R^{M \times M}$ are $\F_t$-measurable and Lipschitz, and correspond to the drift and diffusion respectively. We do not, however, assume any specific parametric form for the drift and diffusion processes but rather learn them via a neural network. The structure of the neural network given in Figure \ref{fig:neural_network_arch} ensures that the drift and diffusion are Lipschitz continuous, as it is essentially a composition of linear transformations and activation functions with Lipschitz constant 1, albeit computing its Lipschitz constant may be challenging. Moreover, the initial condition is fixed, and hence we can guarantee the existence and uniqueness of the solution to the SDE.

At this stage, it is possible to incorporate other time series data, such as equity prices, trading volume, open interest, or any other features the modeler wishes to enhance the data with. This has the added advantage of being able to model the dependence between equity prices, their option's IVs, and other features. For our implementations, we restricted to including only equity prices for simplicity. Moreover, in the multi-asset case, we can concatenate the FPCCs and corresponding market data of each asset to obtain a higher-dimensional time series. We can model this concatenated time series via the neural SDE to capture the correlations across the different assets, as we later show in our results.

\subsection{Neural Architecture} \label{sec:neural_framework}

\begin{figure} [h!]
    \centering
    \includegraphics[width=0.95\columnwidth]{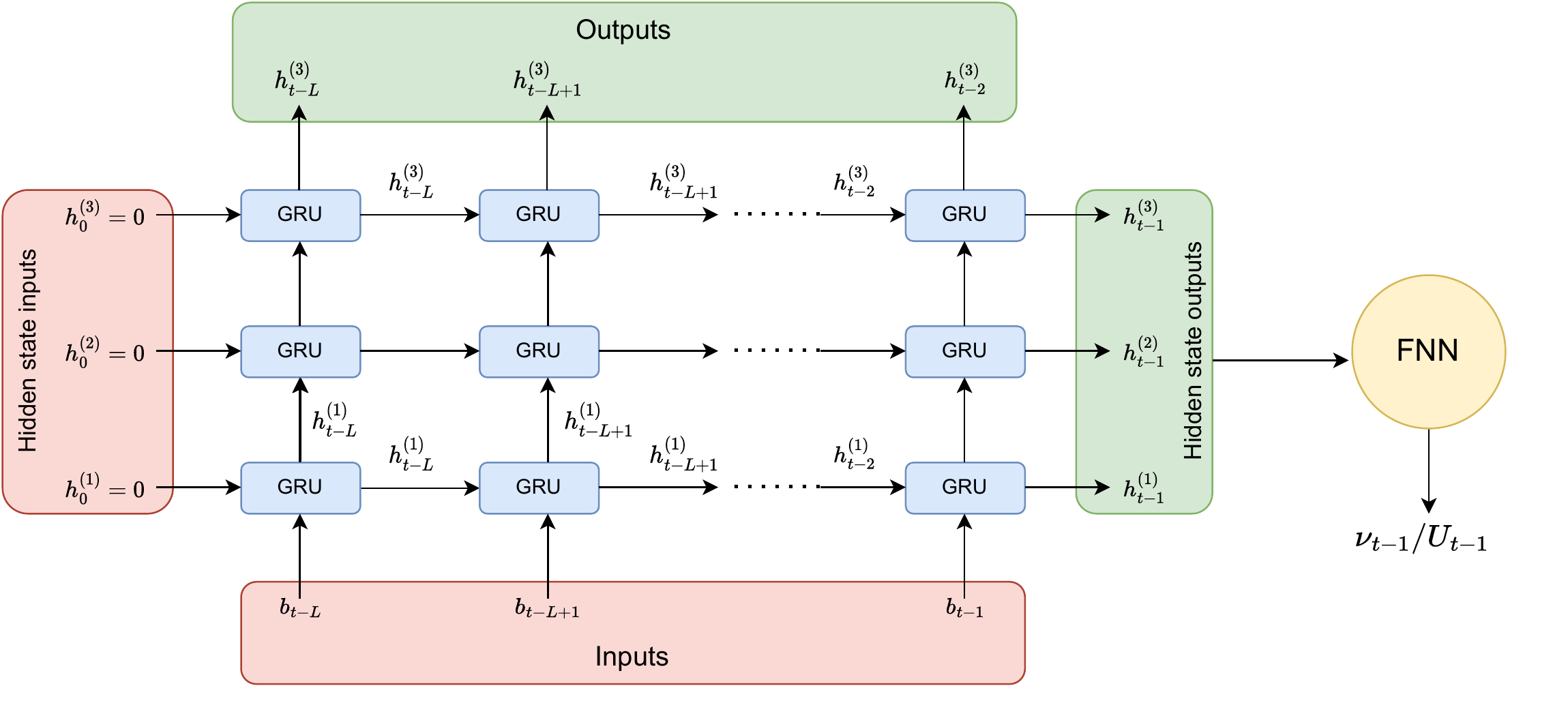}
    \caption{Neural networks architecture for modeling the conditional drift and  diffusion coefficients of the neural SDEs. It consists of 3 stacked layers composed of GRUs spanning across a sequence of length $L$ followed by a feed-forward neural network taking the hidden layer output at the final time as the inputs.} 
    \label{fig:neural_network_arch}
\end{figure}

The SDE in \eqref{eqn:SDE} may be approximated using an Euler discretization by introducing a time grid \{$0,\,\Delta t,\,2\Delta t,\dots$\} and approximating
\begin{equation}
    b_{t+\Delta t} - b_t = \nu_t\ \Delta t + \eta_t\ (W_{t+\Delta t} - W_t)\,.
\end{equation}
As a consequence, we have that
\begin{equation}
    b_{t+\Delta t}|_{\F_t} \sim \N\left(b_t + \nu_t\ \Delta t\,,\; \eta_t\,\eta_t^\intercal \Delta t\right)\,.
\end{equation}
For computational efficiency, we assume the drift and diffusion coefficients depend on the last $L$ lags of the discrete process and are given as outputs of neural networks with parameters $\theta$ and $\gamma$, respectively. Both neural networks have the structure shown in Figure \ref{fig:neural_network_arch}, which consists of multiple gated recurrent unit (GRU) layers (Cho et al. \cite{cho2014learning})  coupled to a feed forward (FF) output layer. Rather than modeling the diffusion directly, the $\gamma$ network models its Cholesky decomposition. We do so, by reshaping the  output of the $\gamma$ network to a lower triangular matrix denoted $L_{t}$ and, to ensure positive definiteness of the resulting covariance matrix,  set $\eta_t\,\eta_t^\intercal =  L_tL_{t}^\intercal+\varepsilon\,I$, where $I$ is the identity matrix and $0<\varepsilon \ll 1$. 

We opt to use GRU layers to learn the path dependency of the FPCCs drift and volatility due to their ability to learn long range dependencies, as well as their ability to tackle the vanishing gradients problem by the use of reset and update gates. 

We perform a neural architecture search with a variety of alternative architectures for the $\theta$ and $\gamma$ neural networks that model the conditional drift and diffusion. For example, instead of having two distinct networks, we experimented with an architecture whereby the outputs of a single three layered GRU were mapped through (a) one single FF neural net,  and (b) two seperate FF neural networks to generate the conditional drifts and diffusions. Both of these architectures under-performed the seperate GRU models and FF layers for drift and diffusion described above. A plausible explanation for this observation is that the non-Markovian features learnt by the hidden states in the GRU are qualitatively different for the conditional drift and conditional diffusion.

\subsection{Training} \label{sec:neural_train}

To enhance the stability of the model, we learn the parameters $\theta$ and $\gamma$ in an iterative manner, and the entire process consists of three steps. 

In the first step, we estimate $\hat{\theta}$ by minimizing the mean squared error (MSE) of the observed time series and its estimated conditional mean. The optimization problem at this stage is given by (here, and in the sequel, the subscript of the FPCCs denote the time count index and not time itself)
\begin{equation}
    \hat{\theta} = \argmin_{\theta}
    \frac{1}{T-L} \sum_{t=L+1}^T (b_t - b_{t-1} - \nu_{t-1} \Delta t)^\intercal (b_t - b_{t-1} - \nu_{t-1} \Delta t)\,.
    \label{eqn:mse-loss}
\end{equation}

In the second step, we learn the parameters $\hat{\gamma}$ of the diffusion neural network while keeping $\hat{\theta}$ fixed at the minima above. In this step, we aim to maximize the log-likelihood of the observed time series. 

To help aid in avoiding model misspecification, we regularize by including an additional penalty in the objective function. In the absence of model misspecification, the sequence of marginal PITs \citep{angus1994probability} has a standard uniform distribution. Hence, we add a penalty corresponding to deviations of the marginal PIT from being uniform across all the features. For completeness, we next recall what the PIT of a sequence of random variables are, followed by a proposition which relates the distribution of the sequence of PITs to the misspecification of the data generating process. Finally, we will define the PIT penalty we work with.

\begin{definition}
    Given a sequence of univariate random variables $\{X_t\}_{t=1,\ldots,T}$ with $X_t$ having the cumulative distribution $F_t$, the probability integral transform is the sequence of random variables $\{U_t\}_{t=1,\ldots,T}$ given by $U_t = F_t(X_t)$.
\end{definition}

\begin{proposition}[\cite{diebold1998evaluating}]\label{prop:pit}
Suppose the true data generating process of univariate $\{X_t\}_{t=1,\ldots,T}$ is governed by the sequence of distributions $\{G_t(X_t|X_{t-1},\ldots,X_1)\}_{t=1,\ldots,T}$ and let the learnt conditional distribution forecasts from the generative model be given by $\{F_t(X_t|X_{t-1},\ldots,X_1)\}_{t=1,\ldots,T}$. In the absence of any model misspecification of the data generating process, in which case the sequence of conditional distributions $\{F_t\}_{t=1,\ldots,T}$ and $\{G_t\}_{t=1,\ldots,T}$ coincide, the sequence of probability integral transforms of $\{X_t\}_{t=1,\ldots,T}$ with respect to $\{F_t\}_{t=1,\ldots,T}$ are i.i.d. Uniform(0,1).
\end{proposition}

Thus, to reduce model misspecification of our generative model, we impose an additional penalty on the sequence of PITs for each feature of the time series. In the case of the time series of FPCCs, we have that $b_{t}|_{\F_t} \sim N\left(\mu_{t-1}, \Sigma_{t-1}\right)$ where $\mu_{t} := b_{t} + \nu_{t}\Delta t$ and $\Sigma_t := \left(L_{t} L_{t}^\intercal+ \epsilon I\right)\Delta t$. For the time series of FPCCs, we obtain a sequence of PITs $(U_t^i)_{\tT, \iM}$ given by $U_t^i = \Phi \left(\frac{b_t^i - (\mu_{t-1})^i}{(\Sigma_{t-1})^{ii}}\right)$, where $\Phi$ is the CDF of the standard normal distribution. 

\begin{definition}
    The PIT penalty corresponding to a sequence $(b_t)_\tT$ is given by
    \begin{equation}
        PIT(b) = \sum_{i=1}^M \int_{0}^1 (\rho_i(u)-1)^2 du\,,
    \end{equation}
    where, for each $\iM$, $\rho_i$ is the Kernel density estimator of $(U_t^i)_\tT$ using the Gaussian kernel with $10\%$ of Silverman's estimate \citep{silverman1986density} as the bandwidth.
\end{definition}
\noindent
Using the estimated value of $\hat{\theta}$ in the first step, the optimization problem in the second step is given by 

\begin{equation}
    \hat{\gamma} = \argmin_{\gamma} \left(-\sum_{t=L+1}^T \log\phi(b_t; \mu_{t-1}, \Sigma_{t-1}) + \alpha\ PIT(b) \right)\,,
    \label{eqn:ll-pit-loss1}
\end{equation}
where $\phi(\cdot;\mu,\Sigma)$ is the density of a multivariate normal with mean $\mu$ and covariance $\Sigma$, and $\alpha$ is a hyperparameter -- we discuss how we set it in Section \ref{sec:res_neural_sde}.

In the third step of our optimization procedure, we jointly optimize both $\theta$ and $\gamma$ using the same objective function as in the second step:
\begin{equation}
    \hat{\theta},\; \hat{\gamma} = \argmin_{\theta, \gamma}
    \left(-\sum_{t=L+1}^T \log\phi(b_t; \mu_{t-1}, \Sigma_{t-1}) + \alpha' \, PIT(b) \right)\,,
    \label{eqn:ll-pit-loss2}
\end{equation}
but with a potentially different value for the hyperparameter $\alpha'$.

\noindent
\subsection{Generating Surfaces} \label{sec:neural_gen}
Once the neural SDE model is trained, we run a forward pass through it using the time series of FPCCs for the last $L$ lags to obtain the mean and covariance estimates for the FPCC at the next time step. We then recursively apply the one-step transition
\begin{equation}
    b_{t+1}|_{\F_t} \sim \N\left(b_t + \nu_t\Delta t\,,\; \Sigma_t \right)\,,
\end{equation}
by sampling from a multivariate normal, and where $\nu_t$ and $\Sigma_t$ are provided by the neural net architecture discussed previously, and use the last $L$ lags of $b$ as inputs. Algorithm \ref{alg:gen} summarizes the approach.

\begin{algorithm}
\caption{Pseudocode to generate surfaces in the future}\label{alg:gen}
\begin{algorithmic}[1]
\Input FPCCs for the last $L$ lags  $b_{t=t_0-L+1:t_0}$, number of future time steps $n$, learnt neural SDE models $\hat{\theta}$ and $\hat{\gamma}$, learnt FPCs $\{\psi_i\}_{\iM}$, $\varepsilon=10^{-3}$
\Output IV surface $(\sigma_t)_{t=t_0+1:t_0+n}$
\For{$t = t_0: t_0+n-1$}
\State $\nu_{t}, L_{t} \; \leftarrow \; b_{t-L:t}$ using $\hat{\theta}$ and $\hat{\gamma}$ networks
\State $b_{t+1} \sim N\left(b_{t}+\nu_{t} \Delta t, \left(L_{t} L_{t}^\intercal+ \varepsilon I\right)\Delta t\right)$
\State store IV surface $\sigma_{t+1} = \sum_{\iM} b_{t+1,i}\,\psi_i$
\EndFor
\end{algorithmic}
\end{algorithm}

\section{Quantifying Arbitrage}
\label{sec:quant-arb}

The  lack of consensus for assessing the quality of generated samples for time series data in the extant literature makes assessing the quality of generated surfaces challenge. For a model to be useful it needs to be able to replicate the data in the market while remaining arbitrage free. While our trained models have PITs that are very close to uniform, which is one measure of goodness of fit from an academic viewpoint, from a traders perspective, it is important to quantify the amount of arbitrage there is in generated surfaces. To this end, in this section, we present metrics for quantifying the amount of arbitrage observed in both the historical data and in the generated surfaces, and we demonstrate that our generated surfaces are in line with historical data in Section \ref{sec:res_sim_surface}. Thus, we provide both the academic and practitioner's evaluation of the model.

Specifically, we focus on butterfly and calendar spread metrics which quantify the presence/absence of static arbitrage and are measures that are often employed in practice for making trading decisions. For this purpose, we use the conditions presented by \cite{gatheral2014arbitrage} (see Lemma 2.1 and 2.2). We then compare the distribution of these metrics for observed and synthetic data over a 30-day period following the training period. To obtain the distribution of the arbitrage metrics for synthetic data, we use 100,000 independent paths with IV generated on the same delta-maturity grid as the observed data.

Although we model the IV as a function of option deltas, Gatheral's conditions on the absence of arbitrage are given in terms of the total implied variance and the log-moneyness $m:=\log \frac{K}{S}$ (where $K$ is the option's strike and $S$ the underlying spot price). Given an option's delta, time to maturity and IV, we  obtain its log-moneyness as follows (assuming zero interest rates):
\begin{equation}
    m = -\Phi^{-1}(\delta) \,\sigma \sqrt{\tau} + \frac{\sigma^2 \tau}{2}\,.
\end{equation}

The total implied variance of an option with log-moneyness $m$ and time to expiry $\tau$ is defined as $w(m,\tau) := \sigma^2(m,\tau)\tau$ where $\sigma(m,\tau)$ is the IV of the option.

\begin{definition}
An IV surface is free of calendar spread arbitrage if 
\begin{equation}
    \partial_{\tau} w(m,\tau) \geq 0, \quad \forall m \in \R, \tau>0\,.
\label{eqn:calendar-spread}
\end{equation}
Correspondingly, we define the calendar spread metrics for a given IV surface by $\partial_{\tau} w(m,\tau)$. Given the IV surface $\sigma_{t}$ on day $t$, the calendar spread metrics are given by\\ $\{CS_{i,j}\}_{i=1,\ldots,n_e-1;  j=1,\ldots,n_d}$ where 
\begin{equation}
    CS_{i,j} = \frac{w(m_j,\tau_i)-w(m_j,\tau_{i-1})}{\tau_i-\tau_{i-1}}\,,
\end{equation}
and $\{m_j\}_{j=1,\dots,n_d}$ and $\{\tau_i\}_{i=1,\dots,n_e}$ are a fixed grid of points on which we estimate the calendar spread.
\end{definition}
\begin{definition}
A slice of the IV surface $w(m)$ (we drop the $\tau$ for simplicity of notation) obtained by fixing maturity $\tau$ is free of butterfly spread arbitrage if 
\begin{equation}
g_\tau(m):=\left(1-\frac{m \partial_{m} w(m)}{2 w(m)}\right)^2-\frac{(\partial_{m} w(m))^2}{4}\left(\frac{1}{w(m)}+\frac{1}{4}\right)+\frac{\partial_{mm}^2 w(m)}{2} \geq 0, \quad \forall m \in \R
\label{eqn:butterfly-spread}
\end{equation}
Correspondingly, we denote the butterfly spread metrics for a given IV $\sigma_t$  on day $t$ by $\{BS_{i,j}\}_{i=1,\ldots,n_e; j=1,\ldots,n_d-2}$ where $BS_{i,j}=g_{\tau_i}(m_j)$ for $\{m_j\}_{j=1,\dots,n_d}$ and $\{\tau_i\}_{i=1,\dots,n_e}$ a given fixed grid of points. The partial derivatives in  \eqref{eqn:butterfly-spread} are estimated by central differencing.
\end{definition}

\section{Results} \label{sec:results}

In this section, we illustrate our methodology to generate synthetic IV surfaces for equity options. In addition to generating IV surfaces, we also generate price paths for the assets. Moreover, we generate synthetic samples for four (4) equities simultaneously. This has the added benefit that each generated scenario is consistent across all four equities. The data used for these purposes along with the relevant transformations are discussed in Section \ref{sec:res_data_trans}. Section \ref{sec:res_data_proj} deals with the functional projection data to obtain the FPCCs, and Section \ref{sec:res_neural_sde} discusses the neural SDE model for FPCCs. Section \ref{sec:res_sim_surface} presents samples of the generated surfaces, as well as price paths, along with the metrics that quantify any arbitrage opportunities found in the generated surfaces. Finally, in Section \ref{sec:res_delta_hedge} we conduct a simple delta-hedging exercise as an additional validation of the quality of the generated data.

\subsection{Data and Transformations}
\label{sec:res_data_trans}

We source IV data for American call options on Amazon (AMZN), Intel (INTC), IBM (IBM) and Tesla (TSLA) and the corresponding asset price for the time period July 8, 2010 to December 31, 2021 from Option Metrics via Wharton Research Data Services [\citenum{wrds}] (WRDS). Data is retained for just those days where all four (4) equities are available, totaling to 2,893 observations. For each ticker, we have IV data on a grid of 17 deltas in the range of 0.1 to 0.9 at intervals of 0.05, and 11 maturities ranging from 10 calendar days to 2 years. The volatility values on this grid are obtained by interpolating actual values, which is done internally in the Option Metrics data, and not by ourselves. Hence, there is no guarantee of the IV surfaces being arbitrage-free. Given that the observed values are on a discrete grid, we first project this data onto the orthonormal Legendre polynomial basis functions to obtain entire surfaces and then perform FPCA to reduce the dimensionality of the problem, as described in Sections \ref{sec:meth_basis_proj} and \ref{sec:meth_obtain_fpc}. The range of values for delta and maturity, however, are not in $[-1,1]$, but rather in $[0,1]\times [0,2]$. The Legendre polynomials are not orthogonal in this range. Therefore, we perform a series of transformations to render them amenable to analysis. 

Firstly, the deltas are transformed as $\Delta \mapsto \tdelta:=2\,\Delta-1$ so that they lie in the range $[-1,1]$. The option maturity are transformed as $\tau \mapsto \ttau := \sqrt{\tau}$ to prevent clustering of observed data near the short maturity axis in the transformed space. This leads to more robust fits of the IV surface towards the longer expiry. This transformation is followed by scaling and shifting the transformed values for them to lie in $[-1,1]$ so that all together $\tau \mapsto \ttau := 2(\sqrt{\tau/ \tau_{n_e}})-1$, where $\tau_{n_e}=2$ is the longest time to expiry in the data. To ensure the positivity of the generated IV surfaces, we perform the transformation  $\sigma\mapsto \tsigma:=c_0+c_1\log(e^\sigma-1)$, where $c_{0,1}$ are constants chosen such that only $10\%$ of the transformed data lies outside the range $[-1,1]$, i.e., $\P(\tsigma<-1)=\P(\tsigma>1)=0.1$. The constants $c_{0,1}$ vary across equities, as the various equities have varying levels of IV. For a particular equity $e$, denote the $q^{th}$ empirical quantile of 
$\{\log(e^{\sigma_{t,i}}-1)\}_{\tT,\iI}$ by $Q_q^e$, then 
\begin{equation}
    c_0^e = -\frac{Q_{0.9}^e+ Q_{0.1}^e}{Q_{0.9}^e - Q_{0.1}^e} \quad \text{and} \quad c_1^e = \frac{2}{Q_{0.9}^e- Q_{0.1}^e}\,.
\end{equation} 
For equity prices, we perform a similar linear transform to ensure the transformed data lie with 90\% probability in the range $[-1,1]$, i.e., $S\mapsto \widetilde{S}:=c_0+c_1\,S$, where $c_{0,1}$ are obtained as above, but using the empirical quantiles of $\{S_t\}_{\tT}$. As with the original prices, the transformed prices have an increasing trend over time, therefore we detrend them by regressing $\beta_0^e + \beta_1^e \,t$ and subtracting the regression from $\widetilde{S}_t$. 

There is a two-fold rationale behind the set transformations on IV surfaces and equity prices. First, the IV transformations ensure that the generated IV surfaces (after inverting the transformations) are always positive. Second, the observed data has vastly different ranges for IV and prices, and it also differs across different equities. The linear transformations normalize the data to enhance the learning of the neural SDE.

\subsection{Projecting data onto surfaces}\label{sec:res_data_proj}

\begin{figure} [!ht]
    \centering\includegraphics[width=0.95\textwidth]{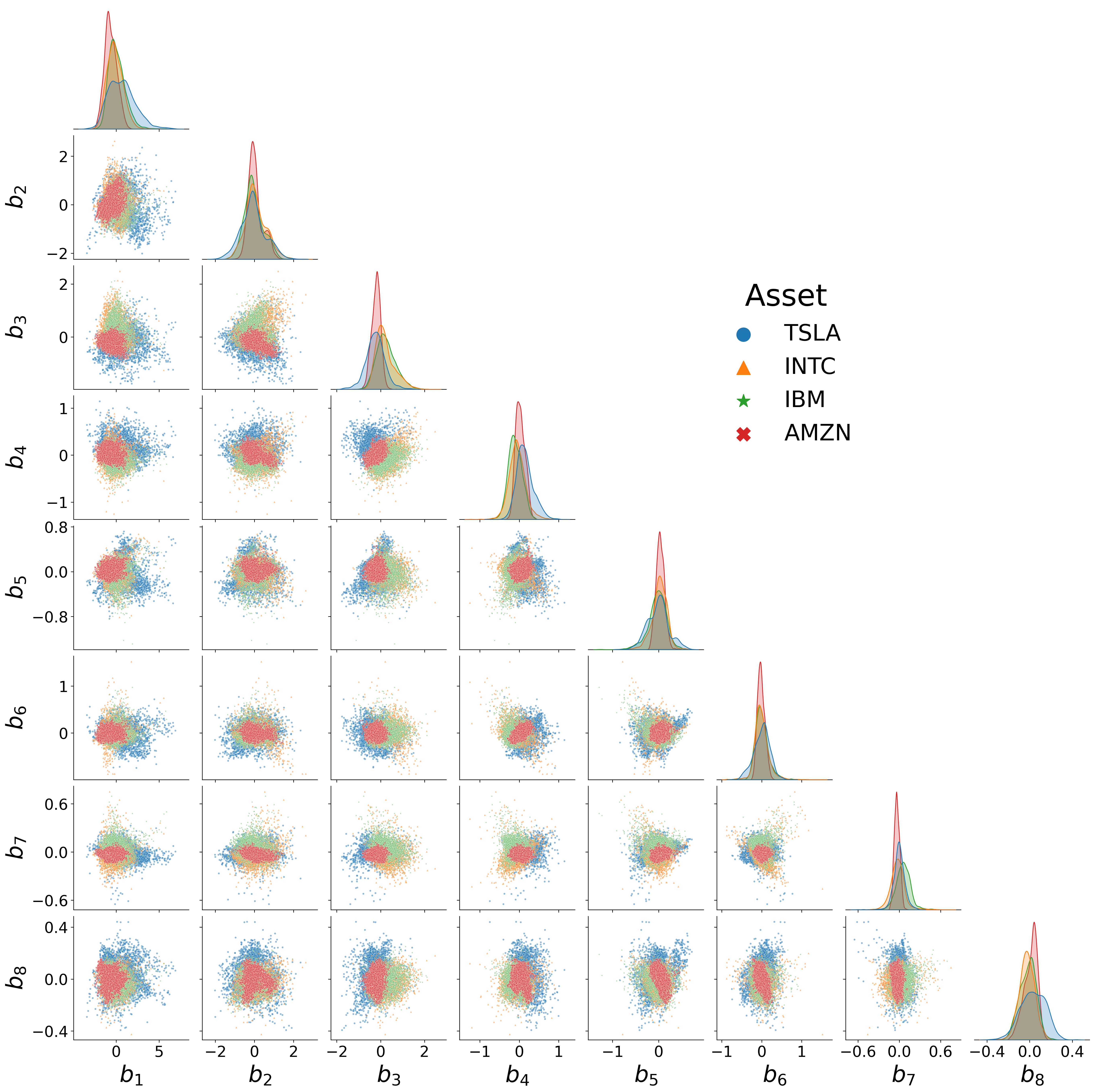}
    \caption{Pairwise scatter plot of the estimated time series of coefficients $\{b_{t,k}\}$ corresponding to the projection of IV data on the functional principal components for all four equities.} 
    \label{fig:pairwise_scatter_b}
\end{figure}

The transformed data is first projected onto orthonormalized Legendre polynomials following the methodology in Section \ref{sec:meth_basis_proj}. Figure \ref{fig:pairwise_scatter_a} shows the pairwise scatter plots of the time series of coefficients $\{a_{t,k}\}$ and suggests there is a high degree of correlation across many pairs. Therefore, we perform FPCA to obtain a collection of functional principal component surfaces to remove this correlation. It is important to note that the FPCs we obtain are common across all assets. Once the FPCs are obtained following Section \ref{sec:meth_obtain_fpc}, we choose the $M$ largest principal components that explain at least $99.5\%$ of the variability in the data, for the given data. We find that $M=8$ FPCs explain more than $99.5\%$ of the variability and they are presented in Figure \ref{fig:fpcs}. 

The shape of the resulting FPCs align well with the expected characteristics of the IV surfaces, namely term structure, skew, and convexity. The first two FPCs highlight the term structure of the IV, with the IV decreasing with increasing time to maturity, the decrease happening at a faster rate in the second FPC. The volatility skew is reflected in the third FPC most prominently in addition to the first two. The fourth FPC appears to induce skew in one direction for the short term and the opposite in the long term -- in others it induces twists and skew reversals in the surface. The fifth FPC mostly accounts for inducing convexity for short term maturities. A combination of all these FPCs enables us to capture the statistical properties of the IV surfaces observed empirically.  

Next, we project the transformed data onto the reduced set of FPCs as per Section \ref{sec:meth_fpc_proj} to obtain the estimated time series of FPCCs $((b_{t,i})_\iM)_{\tT}$. Figure \ref{fig:pairwise_scatter_b} shows the pairwise scatter plots of these coefficients, and in comparison to Figure \ref{fig:pairwise_scatter_a}, we see that there is little to zero pairwise correlation  among these new sets of coefficients -- hence, as expected, the FPCs represent orthogonal degrees of freedom of the surface dynamics. To measure the efficiency of the FPCs in compressing the full IV surface, we compare the RMSE of daily IV surface fits when using the set of Legendre basis functions vs the FPCs. The median RMSE when using the Legendre basis functions vs the FPCs are $0.006, 0.0076$ for AMZN, $0.0063, 0.0072$ for IBM, $0.0079, 0.0092$ for INTC and $0.0082, 0.0103$ for TSLA respectively. Figure \ref{fig:rmse_time} in the appendix shows the evolution of the RMSE over time for all the assets. The figures indicate that  the surface fits are good across time with no period having significantly worse fits when compared to others. Moreover, to have a closer look at the distribution of the RMSE values and how much they differ when compressing or not, we also present histograms of the RMSE in Figure \ref{fig:rmse_hist} in the appendix. The figure indicates that the RMSE when using FPCs is  right shifted very  slightly compared with the full basis, again affirming the fact that the FPCs provide good fits with the benefits of dimension reduction outweighing the slight increase in RMSE.

\subsection{Training the neural SDEs}\label{sec:res_neural_sde}

In this section we describe how the neural SDE is trained. First, we concatenate the time series of the estimated FPCCs and equity prices for all assets, resulting in a 36-dimensional time series. The input time series is normalized feature-wise by subtracting the median followed by division with the inter-quartile range. We next model the transformed time series using the neural SDEs described in Section \ref{sec:neural_sde}. In the sequel, we make a slight abuse of notation and refer to the transformed values by $b_t$ as well. We use the first ninety percent of the available data (2,603 observations from July 8, 2010 to November 5, 2020)  to train the $\theta$ and $\gamma$ networks. We conduct training in the three phases as detailed in Section \ref{sec:neural_train}, and use the AdamW optimizer. In the first step, we perform 700,000 iterations to minimize the mean squared error \eqref{eqn:mse-loss} and obtain an initial estimate $\hat{\theta}$ for the $\theta$ network. The model with the least error is retained and used in the second step where we minimize the sum of the negative log-likelihood and PIT penalty \eqref{eqn:ll-pit-loss1}, with a hyperparameter of $\alpha=1$, by varying only the $\gamma$ network to obtain an estimate $\widehat{\gamma}$. We use 100,000 iterations at this stage. Finally, we minimize the loss \eqref{eqn:ll-pit-loss2}, with a hyperparameter of $\alpha'=100$, over both networks to obtain the final estimates for the $\theta$ and $\gamma$ networks.

The hyperparameter $\alpha'$ is selected as the order of magnitude of the absolute ratio of log-likelihood and the PIT penalty at the end of stage two. With this choice, we give equal weights to both the objectives, and the log-likelihood and the PIT penalty converged individually.  An  alternative we considered was updating $\alpha$ dynamically every 10,000 iterations, with its value being set as the ratio of the log-likelihood and the PIT penalty. This resulted in a less stable training regimen and we opted for the static value mentioned above.

\begin{figure} [!htbp]
    \centering
    \includegraphics[width=0.8\columnwidth]{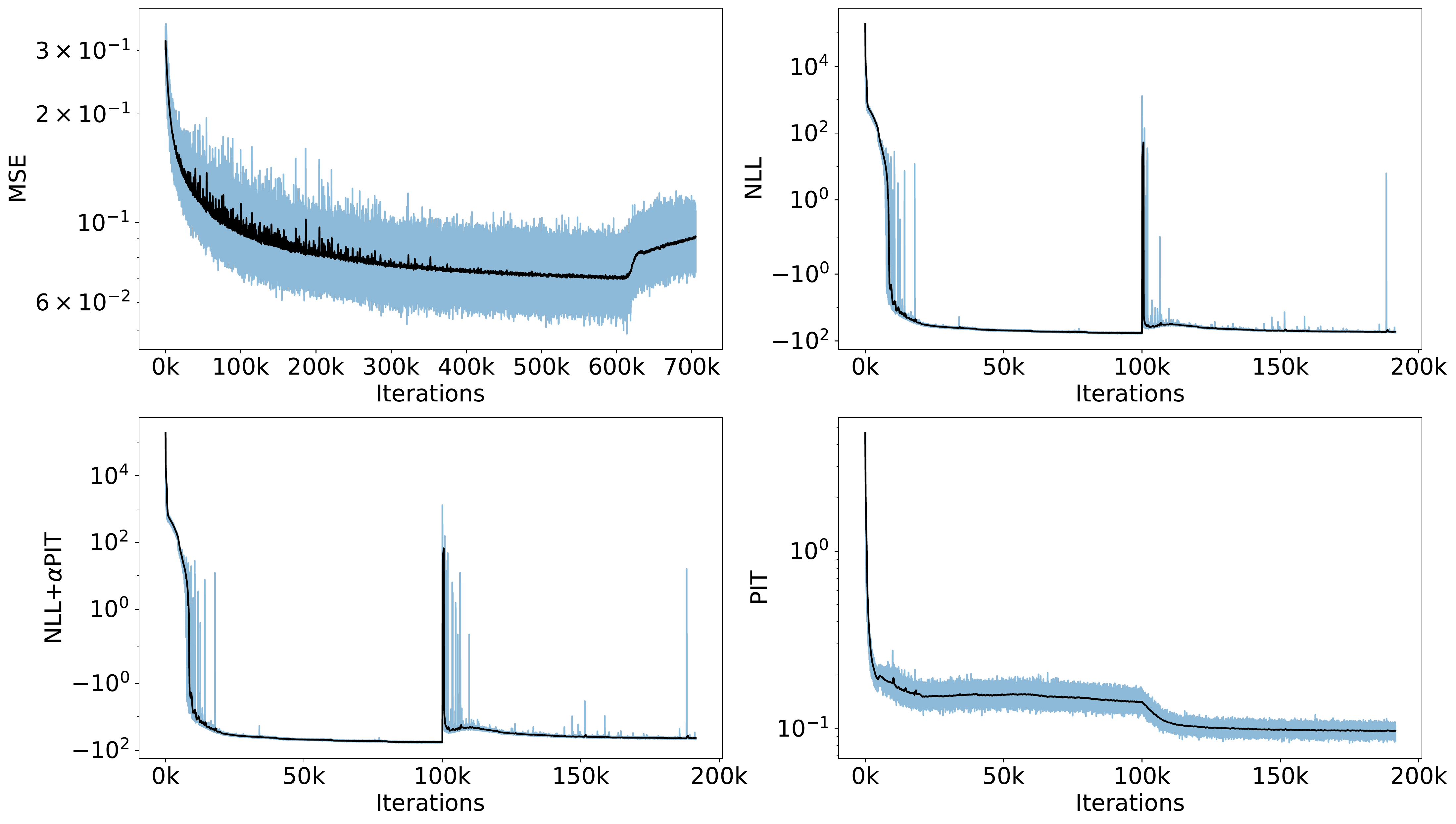}
    \vspace{-6mm}
    \caption{Evolution of the objective functions (in blue) and their moving average over the last 500 iterations (in black) as training progresses. Top: mean squared error (left) and negative log-likelihood (right). Bottom: negative log-likelihood+ $\alpha$ PIT penalty (left), PIT penalty (right)} 
    \label{fig:training loss}
\end{figure}
The various losses as training progresses are shown in Figure \ref{fig:training loss}. All losses converged, with the exception of the MSE. The kink in the MSE where it starts increasing occurs at the onset of the third stage of training where both $\theta$ and $\gamma$ are optimized simultaneously to minimize the sum of negative log-likelihood and the PIT penalty. In this stage, MSE is not explicitly being minimized, hence there is no need for it to converge. Nonetheless, we show the evolution of this loss to assess how far the training of both networks worsens the mean-predictive error. 

\begin{figure} [!htbp]
    \centering
    \includegraphics[width=0.95\columnwidth]{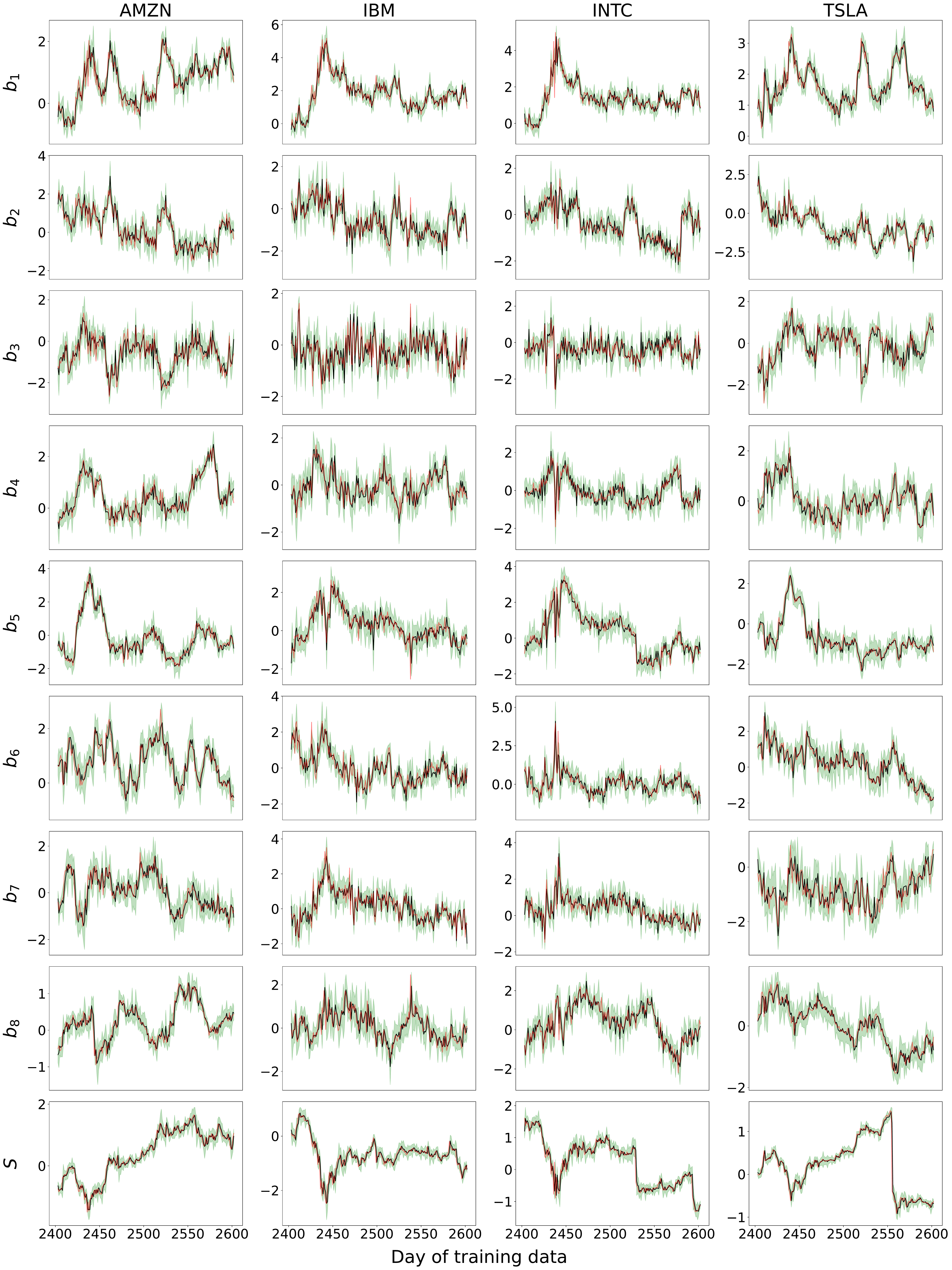}
    \caption{Plot showing the observed transformed FPCCs and prices (in red), its predicted mean (in black) and associated $95\%$ confidence bands (in green) for the last 200 days of training data from the trained model. The first 8 rows correspond to the transformed FPCCs whereas the last row represents the transformed prices. Each column corresponds to one of the equities.} 
    \label{fig:last 200}
\end{figure}
\begin{figure} [!ht]
    \centering
    \includegraphics[width=0.95\columnwidth]{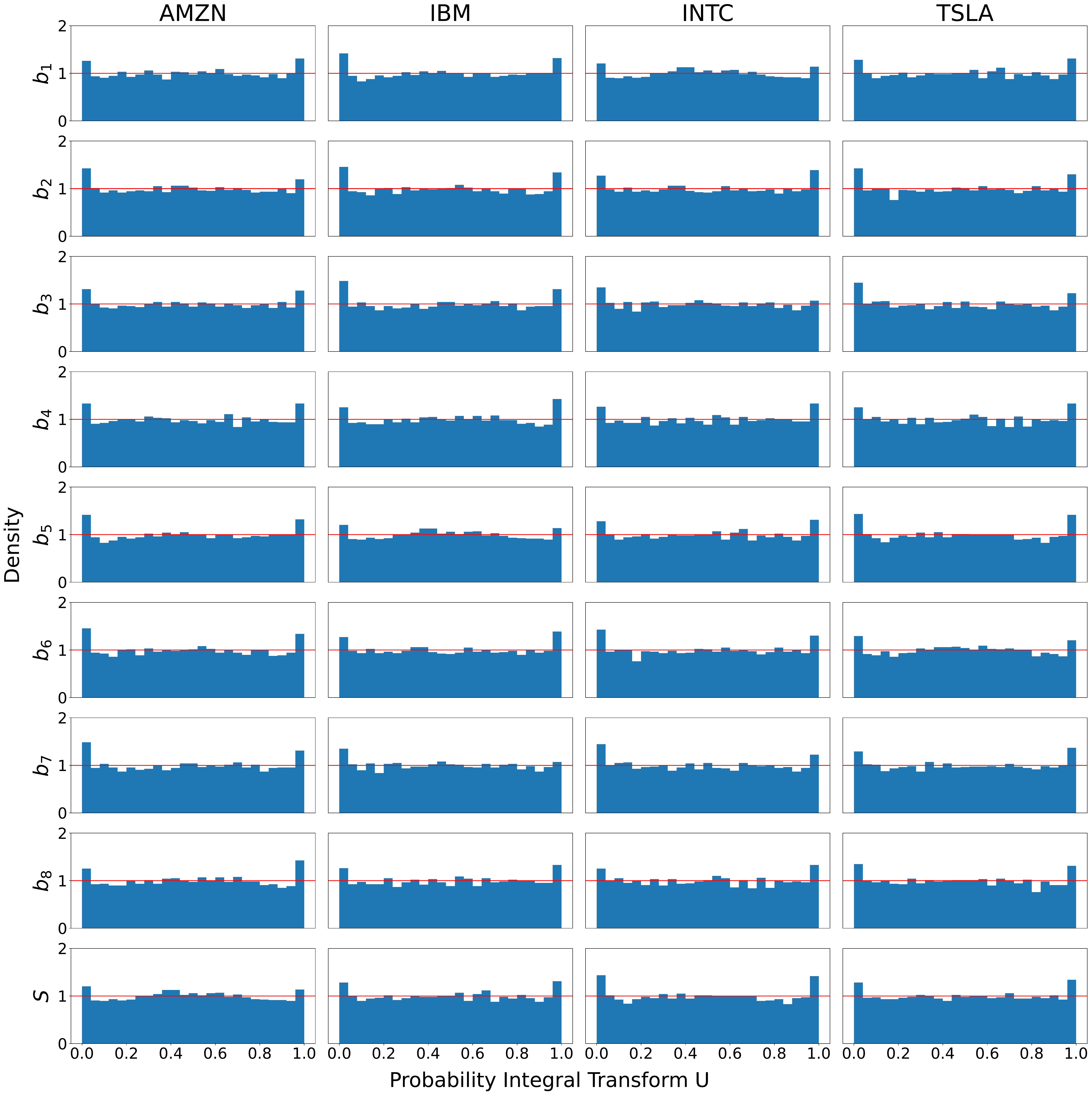}
    \caption{Distribution of the transformed FPCCs and equity prices' PITs at the end of training when PIT penalty is part of objective function.} 
    \label{fig:PIT}
\end{figure}

We retain the model that achieves the minimum loss during stage three of the training for synthetic surface generation. Figure \ref{fig:last 200} shows the predicted mean and $95\%$ confidence bands of the time series of FPCCs and equity prices for all the equities during the last 200 days of training data. The figure suggests that the neural SDE learns the dynamics of the evolution of the time series with the predicted mean closely tracking the observed values and the observed values lying within the confidence bands. The distribution of the PITs of each feature of the time series in Figure \ref{fig:PIT} are very close to $U(0,1)$, indicating little to no model misspecification. The one common observation across the PITs is that there are two small spikes at the very extreme left and right (near $0$ and $1$). These small spikes indicate that the predictive model induced by the neural SDE has just slightly too little weight in the tails compared to the data. One reason for this may is that we use Brownian motions to drive the neural SDE. It may be possible to improve on this by adding in a jump component, such as one driven by a Poisson random measure with compensator given by the output of yet a third neural network. However, as these tail deviations are insignificant, our model fit is acceptable for applications. 

To further demonstrate the impact that the PIT penalty has on reducing model misspecification, we also present the distribution of the PITs when the PIT penalty is not part of the training objective in Figure \ref{fig:PIT_not_in_obj}. The PITs in this case are not as close in distribution to $U(0,1)$ compared to when the PIT was part of the objective function. This is further evidenced by the Kolmogorov-Smirnov test on the PITs. When the PIT penalty is minimized, the null hypothesis is rejected for none of the 36 features in the time series at both the $1\%$ and the $5\%$ significance level whereas in the other case, it is rejected 13 and 25 times respectively. This clearly indicates that including the PIT penalty does indeed reduce model misspecification significantly. Further testing to illustrate that the learnt neural SDE has indeed captured the correlation structure across features as well as the long-range dependencies in the time series are presented in  Appendix \ref{sec:appendix}.

\clearpage
\subsection{Simulated Surfaces}\label{sec:res_sim_surface}
\label{sec:generated-data}

The trained neural SDE model may be used to generate a sequence of FPCCs and prices over multiple consecutive days. These FPCCs then induce IV surfaces through the representation in \eqref{eqn:sigma-from-psi}, and, hence, the simulation provides us with a distribution over sequences of surfaces.  
First, however, we must invert the transformations from Section \ref{sec:res_data_trans}. For our experiments, we generate 100,000 independent paths of $b_{t_e+1:t_e+30}$, where $t_e$ is the end of the training period, and use them to reconstruct surfaces and price paths.

\begin{figure} [!htbp]
    \centering
    \includegraphics[width=0.9\columnwidth]{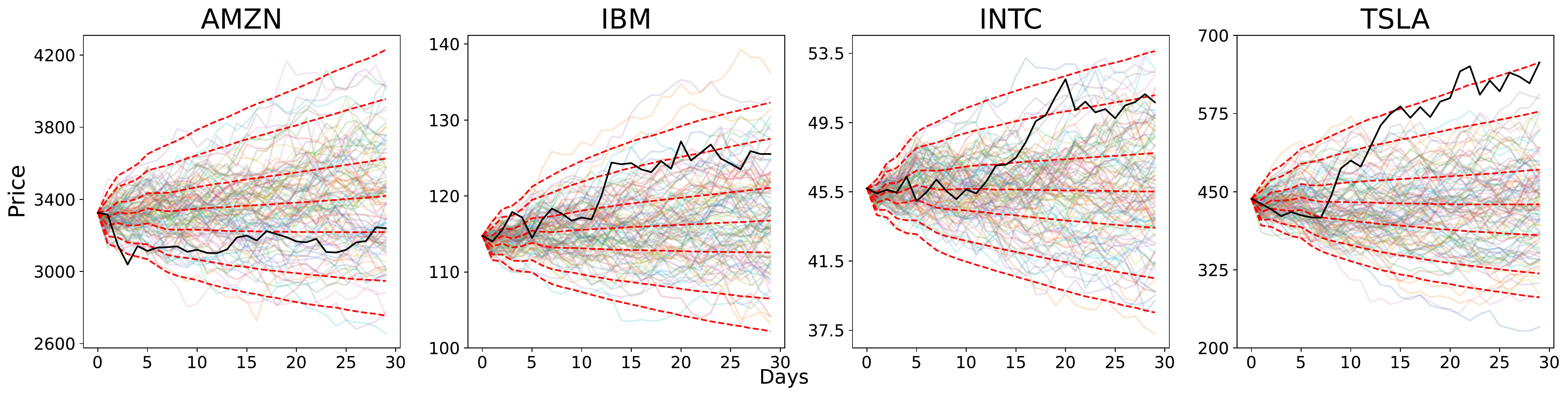}
    \caption{100 generated price paths for the four equities over a 30 day period (November 6 to December 18, 2020) post the training period (July 8, 2010 to November 5, 2020). The actual observed price in black and the quantiles $1^{st}, 5^{th}, 25^{th}, 50^{th}, 75^{th}$ and $95^{th}$ in red.} 
    \label{fig:price quantiles}
\end{figure}
Figure \ref{fig:price quantiles} shows 100 randomly selected (from 100,000) price paths over the 30 days along with the price quantiles (red). By comparing the quantiles with the observed prices (black), the figure suggests that the generated paths are reasonable and can account for scenarios where a drastic upturn or downturn is observed in the market -- e.g., the case of TSLA where the $99^{th}$ quantile closely approximates the actual prices day 15 onwards. The synthetic data is therefore robust for downstream use as it is able to capture extreme scenarios well. 

\begin{figure} [!htbp]
    \centering
    \includegraphics[width=0.9\columnwidth]{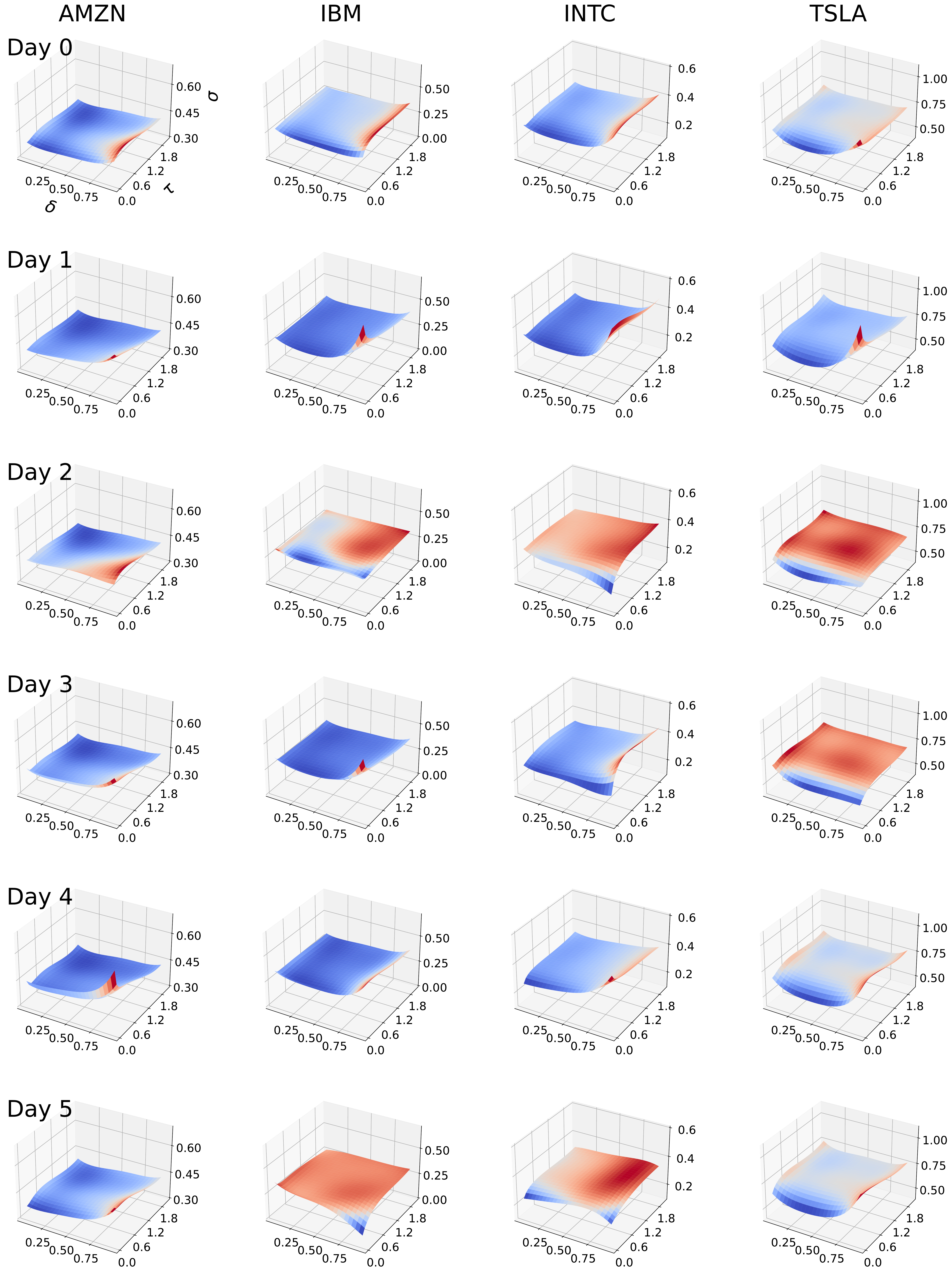}
    \caption{Evolution of the IV surfaces over a 5 day period November 6-12, 2020 (denoted by Day 1-5) for the four equities in one of the scenarios. Day 0 corresponding to November 5, 2020 shows the last of the 10 observed surfaces which are fed as inputs to the model before generation begins.} 
    \label{fig:surface}
\end{figure}
Figure \ref{fig:surface} shows the evolution of the IV surfaces, for the four assets, over a 5 day window for one of the randomly chosen scenarios. The figure illustrates that the generated surfaces show typical dynamics seen in the training data and, as we simultaneously generate all IV surfaces across all assets, we see some dependence across the assets.  Providing confidence intervals across surfaces, and through time, and comparing with the out-of-sample surfaces, akin to what we show for prices in Figure \ref{fig:price quantiles}, is not feasible. Instead, next, we provide a comparison of the calendar and butterfly spread metrics introduced in Section \ref{sec:quant-arb} from our simulated surfaces and compare to the in- and out-of-sample surfaces.

\begin{figure} [!h]
    \centering
    \includegraphics[width=\columnwidth]{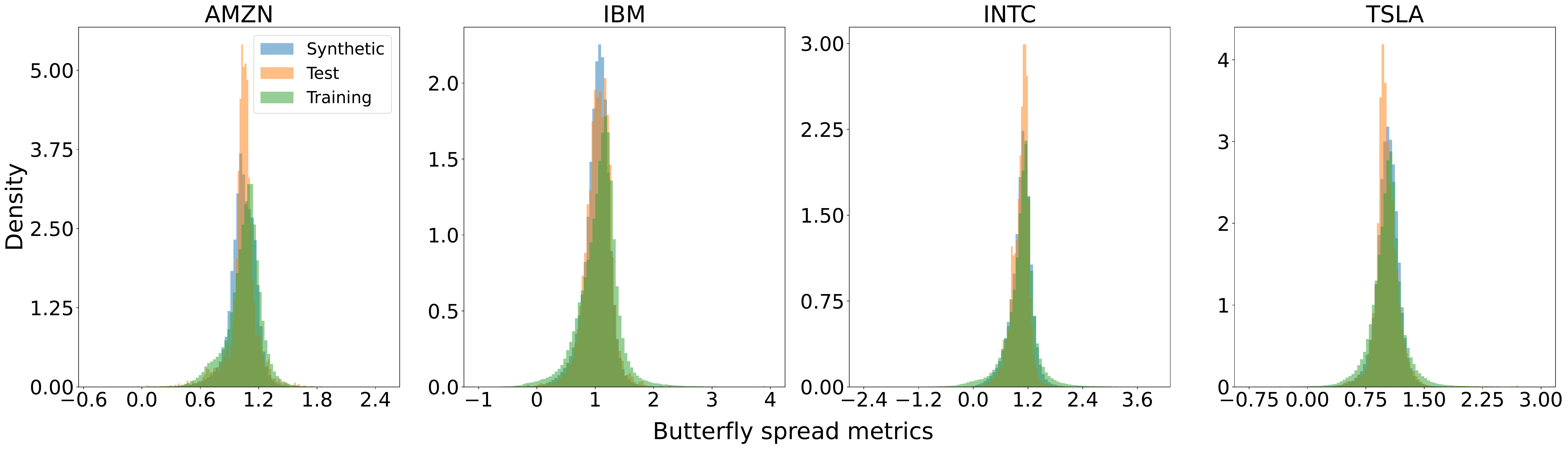}
    \vspace{-6mm}
    \caption{Distribution of the butterfly spread metrics for the training data between July 8, 2010 to November 5, 2020 (Days 1-2603), test data between November 6 to December 18, 2020 (Days 2604-2633) and the 100,000 paths of the generated data for 30 days starting November 6, 2020.} 
    \label{fig:butterfly_spreads}
\end{figure}

\begin{figure} [!h]
    \centering
    \includegraphics[width=\columnwidth]{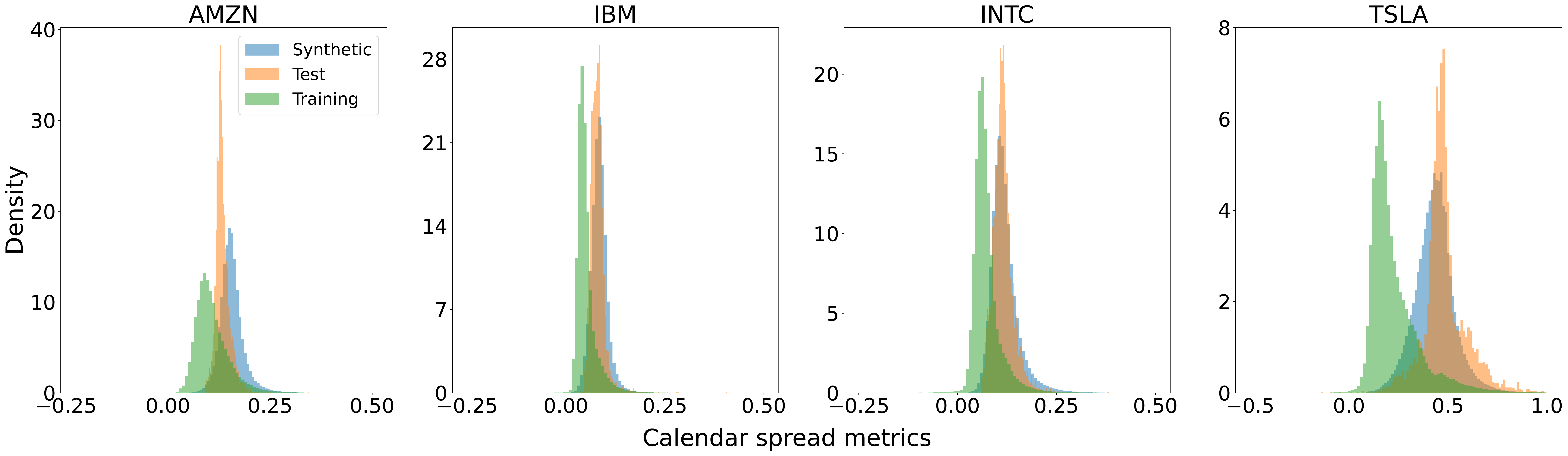}
    \vspace{-6mm}
    \caption{Distribution of the calendar spread metrics for the training data between July 8, 2010 to November 5, 2020 (Days 1-2603), test data between November 6 to December 18, 2020 (Days 2604-2633) and the 100,000 paths of the generated data for 30 days starting November 6, 2020.} 
    \label{fig:calendar_spreads}
\end{figure}

The distribution of the arbitrage metrics for training, test, and synthetic data are shown in Figures \ref{fig:butterfly_spreads} and \ref{fig:calendar_spreads}. The distribution of the butterfly spread metrics overlap to a great extent, however, the training data has slightly heavier tails. The calendar spread metrics for the generated data lie closer to the test data than the training data, and overlaps almost completely for IBM and INTC. The generated (and test) data  are further to the right from zero compared to the training data. This supports the conclusion that the generated surfaces has no more static arbitrage than the training data used to learn the model. As the generated data are simulated from the end of the training period, the local information at the end of this time frame embedded in the hidden states of the GRUs allows our simulated surfaces to capture the shift in the arbitrage metrics from training to test data.

\begin{table}[!h]
\centering
\begin{tabular}{m{0.1\columnwidth}>{\centering}*{8}{m{0.07\columnwidth}}}
 \toprule
  Quantile &  \multicolumn{2}{c}{AMZN}  &  \multicolumn{2}{c}{IBM} &  \multicolumn{2}{c}{INTC} &  \multicolumn{2}{c}{TSLA}\\
  & Obs & Sim & Obs & Sim & Obs & Sim & Obs & Sim\\ [0.5ex]
 \midrule
0.001 & 0.16 & 0.37 & {\color{red}-0.82} & {\color{red}-3.68} & {\color{red}-1.27} & {\color{red}-2.11} & {\color{red}-0.42} & 0.2 \\
0.01 & 0.51 & 0.61 & {\color{red}-0.01} & 0.28 & {\color{red}-0.19} & 0.27 & 0.38 & 0.57 \\
0.05 & 0.70 & 0.81 & 0.47 & 0.61 & 0.41 & 0.59 & 0.68 & 0.78 \\
0.25 & 0.96 & 0.97 & 0.89 & 0.92 & 0.91 & 0.93 & 0.92 & 0.95 \\
0.50 & 1.07 & 1.04 & 1.10 & 1.05 & 1.09 & 1.08 & 1.04 & 1.03 \\
0.75 & 1.15 & 1.13 & 1.25 & 1.16 & 1.21 & 1.19 & 1.13 & 1.12 \\
0.95 & 1.29 & 1.23 & 1.52 & 1.33 & 1.53 & 1.39 & 1.36 & 1.25 \\
0.99 & 1.44 & 1.33 & 2.05 & 1.52 & 2.13 & 1.61 & 1.72 & 1.39 \\
0.999 & 1.79 & 1.56 & 3.33 & 1.97 & 3.78 & 2.23 & 2.83 & 1.78 \\
 \bottomrule
\end{tabular}
\caption{Quantiles of the butterfly spread metrics for the training data  between July 8, 2010 to November 5, 2020 denoted by Obs and for 100,000 simulated paths of length 30 days post the training period (November 6 to December 18, 2020) denoted by Sim.}
\label{tab:butterfly}
\end{table}

\begin{table}[!h]
\centering
\begin{tabular}{m{0.1\columnwidth}>{\centering}*{8}{m{0.07\columnwidth}}}
 \toprule
  Quantile &  \multicolumn{2}{c}{AMZN}  &  \multicolumn{2}{c}{IBM} &  \multicolumn{2}{c}{INTC} &  \multicolumn{2}{c}{TSLA}\\
  & Obs & Sim & Obs & Sim & Obs & Sim & Obs & Sim\\ [0.5ex]
 \midrule
0.001 & 0.03 & 0.07 & {\color{red}-0.03} & 0.02 & {\color{red}-0.13} & 0.04 & {\color{red}-0.07} & 0.12 \\
0.01 & 0.04 & 0.09 & 0.02 & 0.04 & 0.02 & 0.06 & 0.07 & 0.19 \\
0.05 & 0.06 & 0.11 & 0.02 & 0.05 & 0.04 & 0.08 & 0.1 & 0.27 \\
0.25 & 0.08 & 0.14 & 0.04 & 0.07 & 0.05 & 0.1 & 0.15 & 0.37 \\
0.50 & 0.1 & 0.15 & 0.04 & 0.08 & 0.07 & 0.11 & 0.19 & 0.43 \\
0.75 & 0.13 & 0.17 & 0.06 & 0.09 & 0.08 & 0.13 & 0.28 & 0.49 \\
0.95 & 0.19 & 0.21 & 0.1 & 0.12 & 0.14 & 0.19 & 0.49 & 0.6 \\
0.99 & 0.26 & 0.26 & 0.15 & 0.15 & 0.22 & 0.25 & 0.73 & 0.73 \\
0.999 & 0.36 & 0.36 & 0.3 & 0.22 & 0.46 & 0.4 & 1.29 & 0.94 \\
 \bottomrule
\end{tabular}
\caption{Quantiles of the calendar spread metrics for the training data between July 8, 2010 to November 5, 2020 denoted by Obs and for 100,000 simulated paths of length 30 days post the training period (November 6 to December 18, 2020) denoted by Sim.}
\label{tab:calendar}
\end{table}

To further quantify arbitrage, we show in Tables \ref{tab:butterfly} and \ref{tab:calendar} the quantiles of the butterfly and calendar spread metrics for the training data and synthetic data (over 100,000 scenarios) for all four equities. The generated data has positive butterfly spread metrics except for IBM and INTC at the $0.001$ quantile. This is in line with the training data, which also has negative metrics for in these extreme cases. While for IBM and INTC the training data has negative metrics at the $0.01$ quantile, the synthetic data has positive metric values at these levels which indicates that the generated surfaces are free of static-arbitrage. The results are even more encouraging when considering the calendar spread metrics where our simulated surfaces have positive metrics at all quantile levels across all equities. This extends even to the $0.001$ quantile level where, in the training data, all equities except AMZN have negative calendar spread metrics.

To further quantify arbitrage across an entire surface, we introduce a summary statistic that equals one if any point on a surface has a negative arbitrage metric and zero otherwise. The sum of this summary statistic over all test days and training days are reported in the last two columns of Tables \ref{tab:butterfly_days} and \ref{tab:calendar_days}. To compare with our simulator, we generate the summary statistics along every simulated path of thirty days, and sum them. The first seven columns  of Tables \ref{tab:butterfly_days} and \ref{tab:calendar_days} reports the quantiles of this total summary statistics across the simulation scenarios. Thus, for e.g., looking at Table \ref{tab:butterfly_days}, for IBM at the 95\% quantile there are six out of thirty days that have at least one point on the IV surface that contains a butterfly spread arbitrage, for the test data there are three days in the same period, while in the training data that number is 1,457 days out of 2,603 days. Note that to compute the metrics using historical data, we use finite difference approximations for computing the derivatives that appear in \eqref{eqn:calendar-spread} and \eqref{eqn:butterfly-spread}. From our simulated data, we could use a much finer grid, or in principle derive semi-analytical formulae for the same metrics, in which case our metrics will improve even further.

\begin{table}[!h]
\centering
\begin{tabular}{m{0.1\columnwidth}>{\centering}*{8}{m{0.05\columnwidth}>{\centering\arraybackslash}}m{0.12\columnwidth}}
 \toprule
  Equity &  \multicolumn{7}{c}{Quantiles}  &  Test &  Training\\
  & 1 & 5 & 25 & 50 & 75 & 95 & 99 & & \\ [0.5ex]
 \midrule
AMZN & 0 & 0 & 0 & 0 & 0 & 0 & 2 & 1 & 115 (2)\\
IBM & 0 & 0 & 0 & 1 & 2 & 6 & 10 & 3 & 1457 (17)\\
INTC & 0 & 0 & 0 & 1 & 2 & 6 & 11 & 2 & 1697 (20)\\
TSLA & 0 & 0 & 0 & 0 & 0 & 3 & 6 & 2 & 347 (4)\\
 \bottomrule
\end{tabular}
\caption{Quantiles of the number of days having at least one negative butterfly spread metric for the generated data between November 6 to December 18, 2020 across the 100,000 scenarios as well as the observed number of days with negative butterfly spread metrics in the test (November 6 to December 18, 2020) and training (July 8, 2010 to November 5, 2020) data. The numbers in brackets for training data indicate the equivalent number of days with negative metrics on a scale of 30 rather than 2603 days.}
\label{tab:butterfly_days}
\end{table}

\begin{table}[!h]
\centering
\begin{tabular}{m{0.1\columnwidth}>{\centering}*{8}{m{0.05\columnwidth}>{\centering\arraybackslash}}m{0.1\columnwidth}}
 \toprule
  Equity &  \multicolumn{7}{c}{Quantiles}  &  Test &  Training\\
  & 1 & 5 & 25 & 50 & 75 & 95 & 99 & & \\ [0.5ex]
 \midrule
AMZN & 0 & 0 & 0 & 0 & 0 & 0 & 0 & 0 & 26 (1)\\
IBM & 0 & 0 & 0 & 0 & 0 & 0 & 0 & 0 & 517 (6)\\
INTC & 0 & 0 & 0 & 0 & 0 & 0 & 3 & 0 & 892 (11)\\
TSLA & 0 & 0 & 0 & 0 & 0 & 0 & 0 & 3 & 337 (4)\\
 \bottomrule
\end{tabular}
\caption{Quantiles of the number of days having at least one negative calendar spread metric for the generated data between November 6 to December 18, 2020 across the 100,000 scenarios as well as the observed number of days with negative butterfly spread metrics in the test (November 6 to December 18, 2020) and training (July 8, 2010 to November 5, 2020) data. The numbers in brackets for training data indicate the equivalent number of days with negative metrics on a scale of 30 rather than 2603 days.}
\label{tab:calendar_days}
\end{table}

\subsection{Delta Hedging}\label{sec:res_delta_hedge}

To further ensure the accuracy, and faithfulness to historical patterns, of the generated IV surfaces and asset prices, we conduct a delta hedging exercise. This involves implementing a self-financing strategy using the underlying asset and an interest-free bank account. On Day $0$, the last day of our training data (November 5, 2020), we sell an at-the-money (ATM) European call option expiring in $\tau^h_0=30$ days with a strike denoted  $K^h$. At day $t$, the option's time to expiry is now $\tau_t^h:= \tau_0^h - t$, the underlying asset price is $S_t$, the price of the option is $P_t$, and the corresponding IV is $\Delta_t$. We implement a hedge that equals to the prevailing delta in the market, which is the solution to the non-linear equation 
\begin{equation}
    \Delta_t^h = \Phi\left(\frac{\log(S_t/K)+(\sigma_t(\Delta_t^h,\,\tau_t^h)^2/2)\tau_t^h}{\sigma_t(\Delta_t^h,\,\tau_t^h) \sqrt{\tau_t^h}}\right)\,
\end{equation}
where $\sigma_t(\cdot,\cdot)$ is the simulated IV surface at time $t$.  This delta represents the number of units of asset that the hedge strategy holds at time $t$. 

To be delta neutral, we hedge our position at day 0 by having a long position of $\Delta^h_0$ in the underlying asset, where $\Delta^h_t$ represents the Black-Scholes delta of the sold European call at day $t$. Here we make two simplifying assumptions: (i) zero-interest rates, an assumption we make throughout the paper, and (ii) zero transaction costs with an ability to purchase fractional units of the asset. As a result, the bank account has a balance of $B_0 = P_0 - \Delta^h_0 S_0$. Subsequently, we rebalance our position every day to stay delta neutral by buying $(\Delta^h_t-\Delta^h_{t-1})$ units of the underlying so that the bank account at the end of day $t$ is given by $B_t = B_{t-1} - (\Delta^h_t-\Delta^h_{t-1}) S_t$. Finally, at day $\tau_0^h$ we settle our position by paying the option payoff $(S_{\tau_0^h}-K)_+$ as well as liquidating our position in the underlying leaving us with a P\&L of $B_{\tau_0^h-1} - (S_{\tau_0^h}-K)_+ + \Delta^h_{\tau_0^h-1} S_{\tau_0^h}$. 
\begin{figure} [t]
    \centering
    \includegraphics[width=\columnwidth]{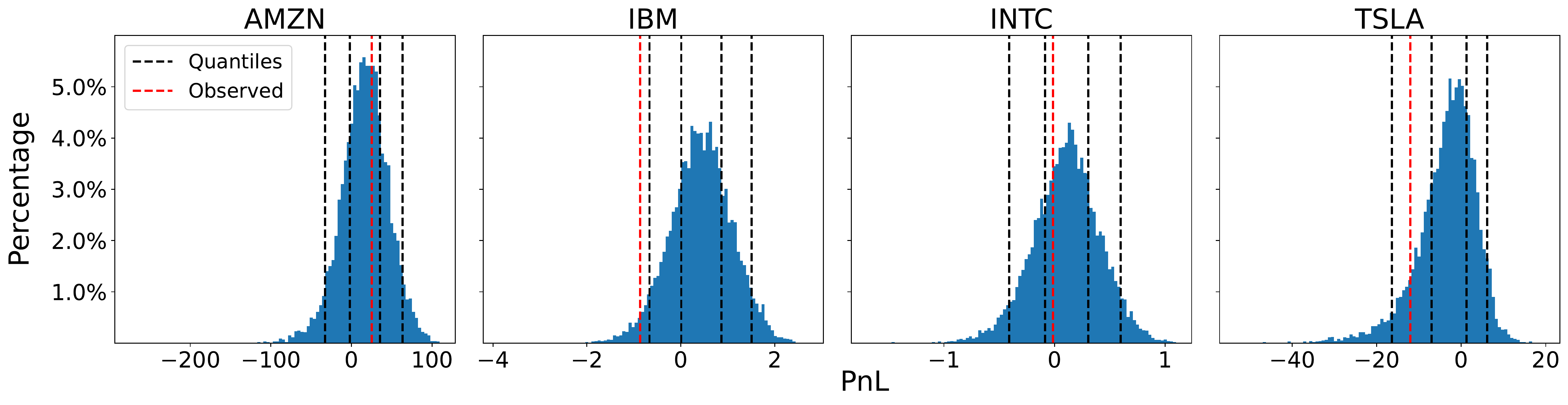}
    \vspace{-2em}
    \caption{Distribution of the P\&L when delta-hedging across 10,000 independent scenarios of the generated data on a 30-day horizon. The red dashed line denotes the P\&L corresponding to realised  data whereas the black dashed lines correspond to the $5^{th}, 25^{th}, 75^{th}$ and $95^{th}$ quantiles.} 
    \label{fig:PnL}
\end{figure}

We execute the delta hedging procedure on four equities using 10,000 simulated paths and the actual path. The distribution of the P\&L across the synthetic paths and the observed data is depicted in Figure \ref{fig:PnL}. To establish a reference point for the P\&L, we report the initial price of the at-the-money (ATM) option and the asset on Day 0, as well as other features such as IV and Black-Scholes delta in Table \ref{tab:delta_hedge}. Figure \ref{fig:PnL} indicates that the P\&L distribution is centered around zero and the P\&L generated from the realized path lands in the bulk of this distribution. Additionally, considering the ATM option's price and the asset price scale, we deem the range of P\&L appears to be reasonable.
\begin{table}[!h]
\centering
\begin{tabular}{crrrr}
\toprule
\toprule
   &  AMZN  &  IBM &  INTC &  TSLA\\ [0.5ex]
\midrule
IV & 0.37 & 0.24 & 0.29 & 0.59\\
Delta & 0.52 & 0.51 & 0.52 & 0.53\\ 
Option price & 140.97 & 3.14 & 1.51 & 28.87\\  
Asset price & 3,325.97 & 114.75 & 45.69 & 438.64\\
P\&L observed & 25.0 & -0.87 & -0.01 & -12.08\\
\bottomrule
\bottomrule
\end{tabular}
\caption{Implied volatility, Black-Scholes delta, and option price of the ATM European call option sold at Day 0 as well as the corresponding asset price on that day. Final row denotes the P\&L obtained following this strategy when working with market observed data.}
\label{tab:delta_hedge}
\end{table}

\section{Conclusions}

In this paper we developed a combined functional data projection approach, coupled with neural SDEs, to model dynamical surfaces. The approach allows us to faithfully represent the historical dynamics of IV surfaces across multiple assets simultaneously. Our generative model is also able to produce essentially arbitrage free surfaces, even though the training data contains arbitrage and no additional penalties have been added for deviations from the arbitrage free submanifold of surfaces. We are able to generate IV surfaces simultaneously for multiple assets as well as asset prices themselves. If the user wishes, they may add any additional features into the neural SDE modeling to enhance the surface generation. Another avenue for future work is to experiment with the neural architecture for the drift and diffusion by considering, for instance, forward-backward GRU layers, or replace them entirely with long short term memory (LSTM) layers, attention networks, and so on. There are a multitude of potential applications of the resulting surface generator, including but not limited to using the surfaces to generate hedging strategies for path-dependent financial options, such as auto-callables, portfolio allocation problems where options are part of the portfolio, and/or obtaining statistical arbitrage strategies with options.

\acks{S.J. acknowledges the support of the Natural Sciences \& Engineering Research council of Canada through NSERC Alliance [ALLRP 550308 - 20] and the Data Science Institute, University of Toronto.
The authors would also like to thank Ivan Sergienko, Brian Ning, and Nicholas Fung, John Hull, Zisis Poulos, and Jacky Chen for their comments on earlier versions of this work.} 

\section*{Disclosure statement}
The authors report there are no competing interests to declare.

\vskip 0.2in

\bibliography{main}

\appendix 
\section{Additional testing} \label{sec:appendix}
In this section, we provide additional tests to demonstrate the goodness of fit of the IV surfaces using FPCs and to conclude that the learnt neural SDE has indeed captured the correlation structure across features as well as the long-range dependencies present in the observed time series. 

\begin{figure} [!ht]
    \centering\includegraphics[width=0.95\textwidth]{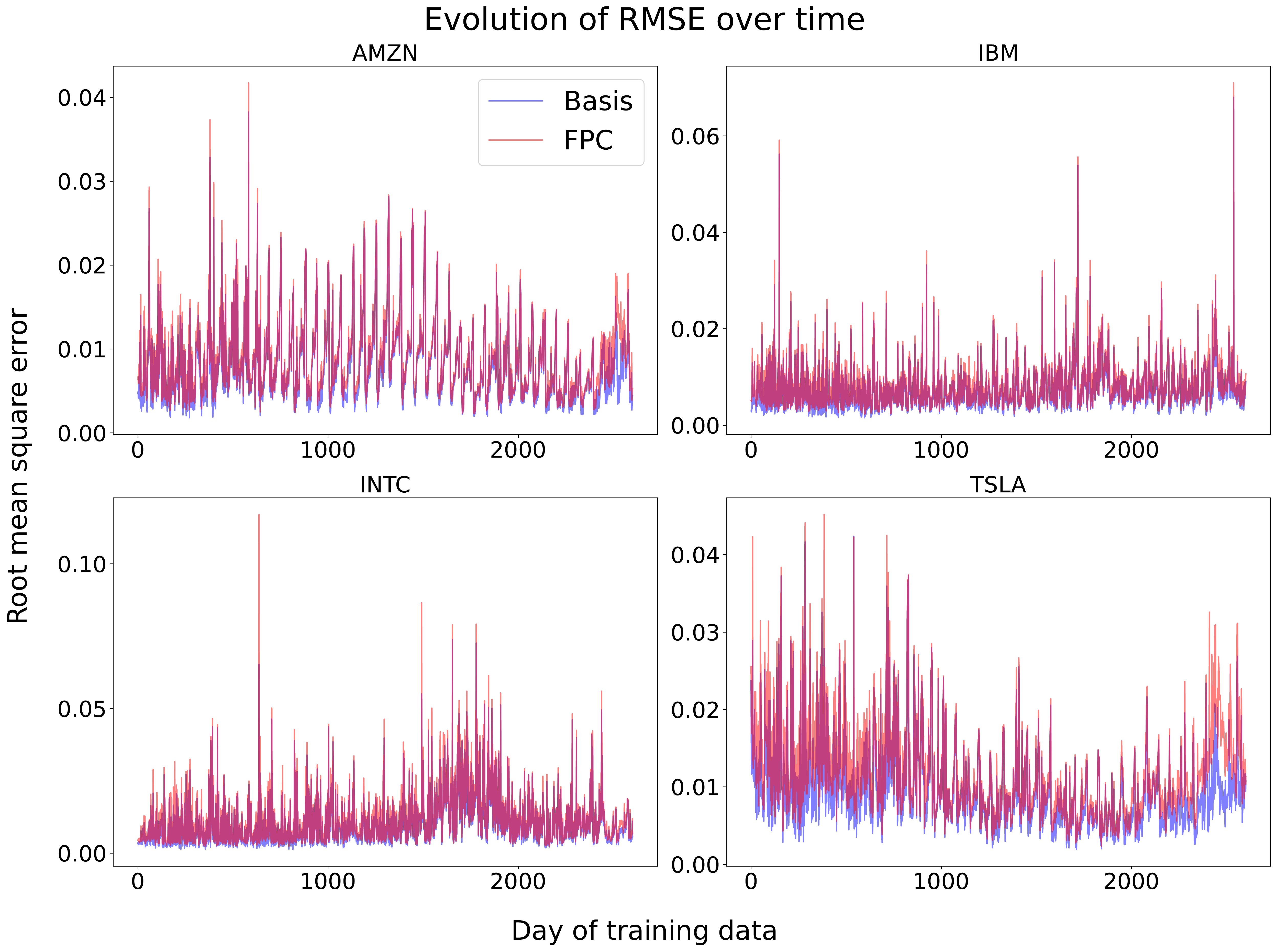}
    \caption{Evolution of the RMSE for the IV surface fits over time when projecting using the basis functions (blue) and the FPCs (red).} 
    \label{fig:rmse_time}
\end{figure}

\begin{figure} [!ht]
    \centering\includegraphics[width=0.95\textwidth]{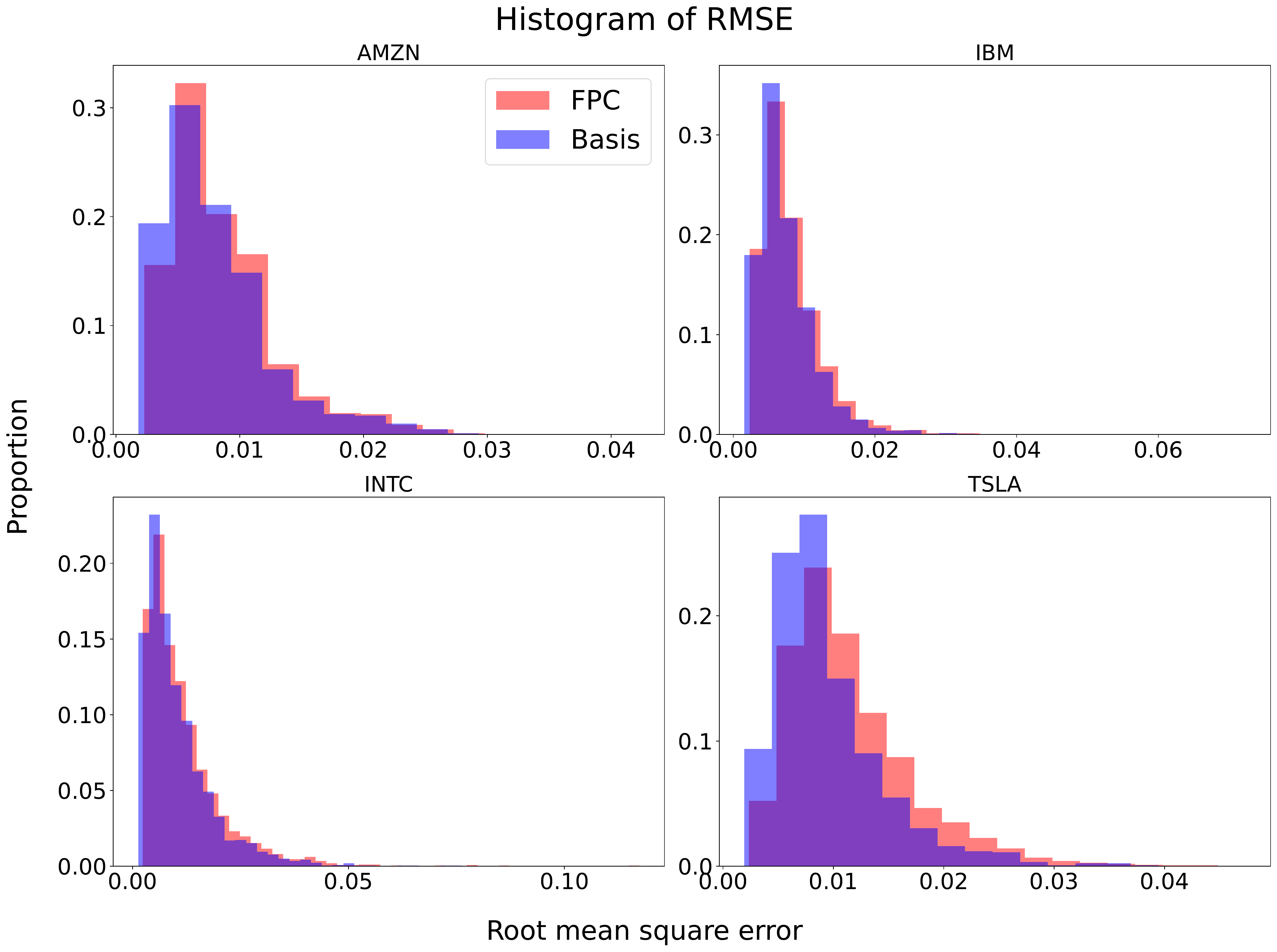}
    \caption{Histogram of the RMSE for the IV surface fits when projecting using the basis functions (blue) and the FPCs (red).} 
    \label{fig:rmse_hist}
\end{figure}

\begin{figure} [!ht]
    \centering
    \includegraphics[width=0.95\columnwidth]{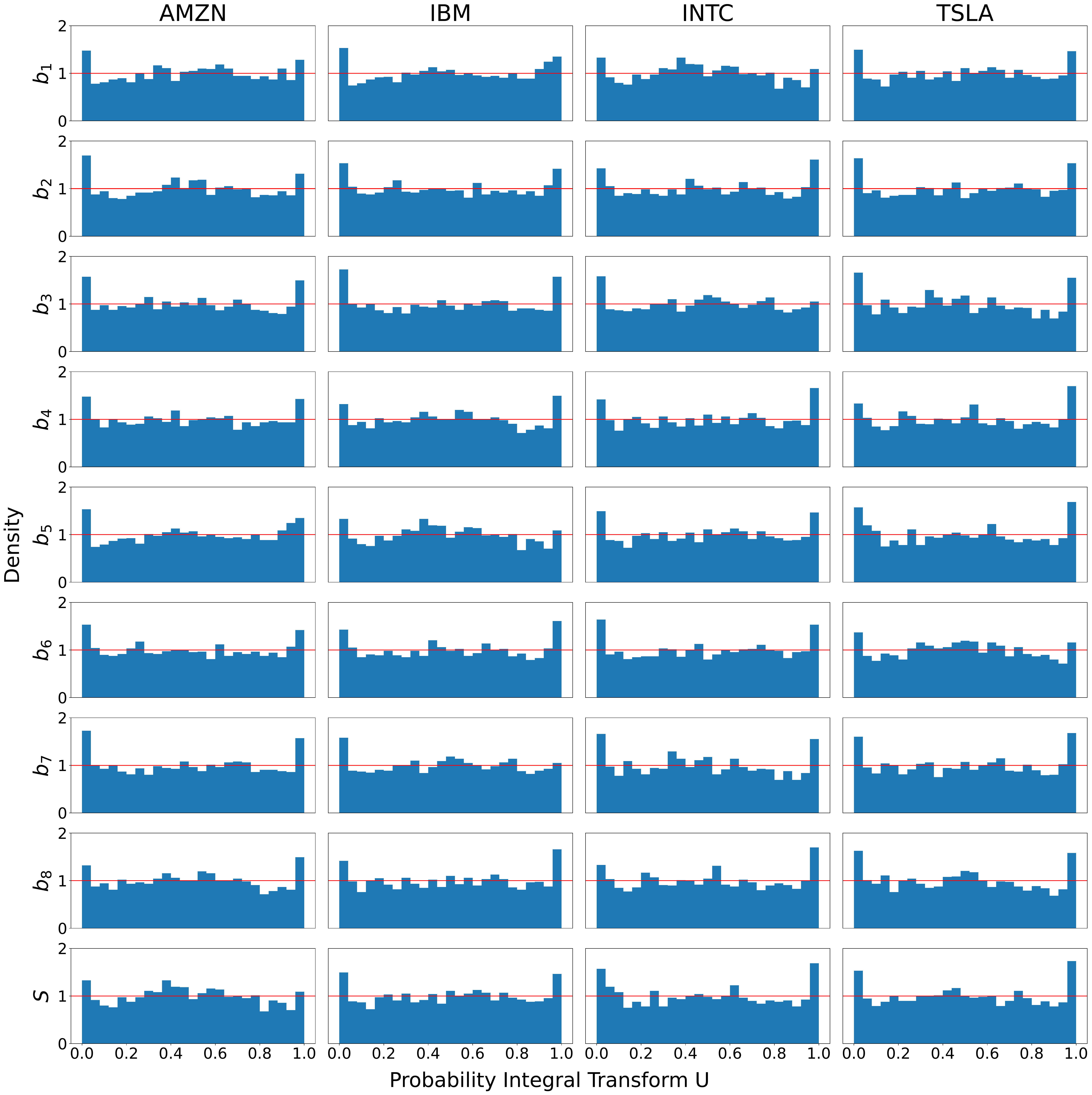}
    \caption{Distribution of the transformed FPCCs and equity prices' PITs at the end of training when PIT penalty is not part of objective function.} 
    \label{fig:PIT_not_in_obj}
\end{figure}

\begin{enumerate}
    \item Auto-correlation: Financial time series are non-stationary and are known to undergo regime changes with different regimes having different auto-correlation structures. Looking at the auto-correlation of the entire time series in such a case can be misleading as it tends to capture the overall average structure. Even if the learned model is able to replicate this overall auto-correlation function (ACF), it does not guarantee that the model has captured the intricacies of the ever-changing ACF. Rather, the learned model may have learnt a stationary time series with the overall ACF of the observed time series. 
    
    Hence, we resort to considering the ACF of the univariate PITs of each feature in the time series. If the ACF of the PITs is zero for all lags, in light of Proposition \ref{prop:pit} we can conclude that the model has captured the ACF structure inherent in the observed non-stationary time series. We have already shown that the PITs are uniformly distributed, and in case the ACF structure was not correctly captured, the PITs will not be i.i.d. $U(0,1)$. The plot for the ACF of the PITs in Figure \ref{fig:PIT_ACF} shows that the ACF is very close to zero most of the time, with only a few features having a slight negative auto-correlation at lag 1. 
    
\begin{figure} [!ht]
    \centering
    \includegraphics[width=0.95\columnwidth]{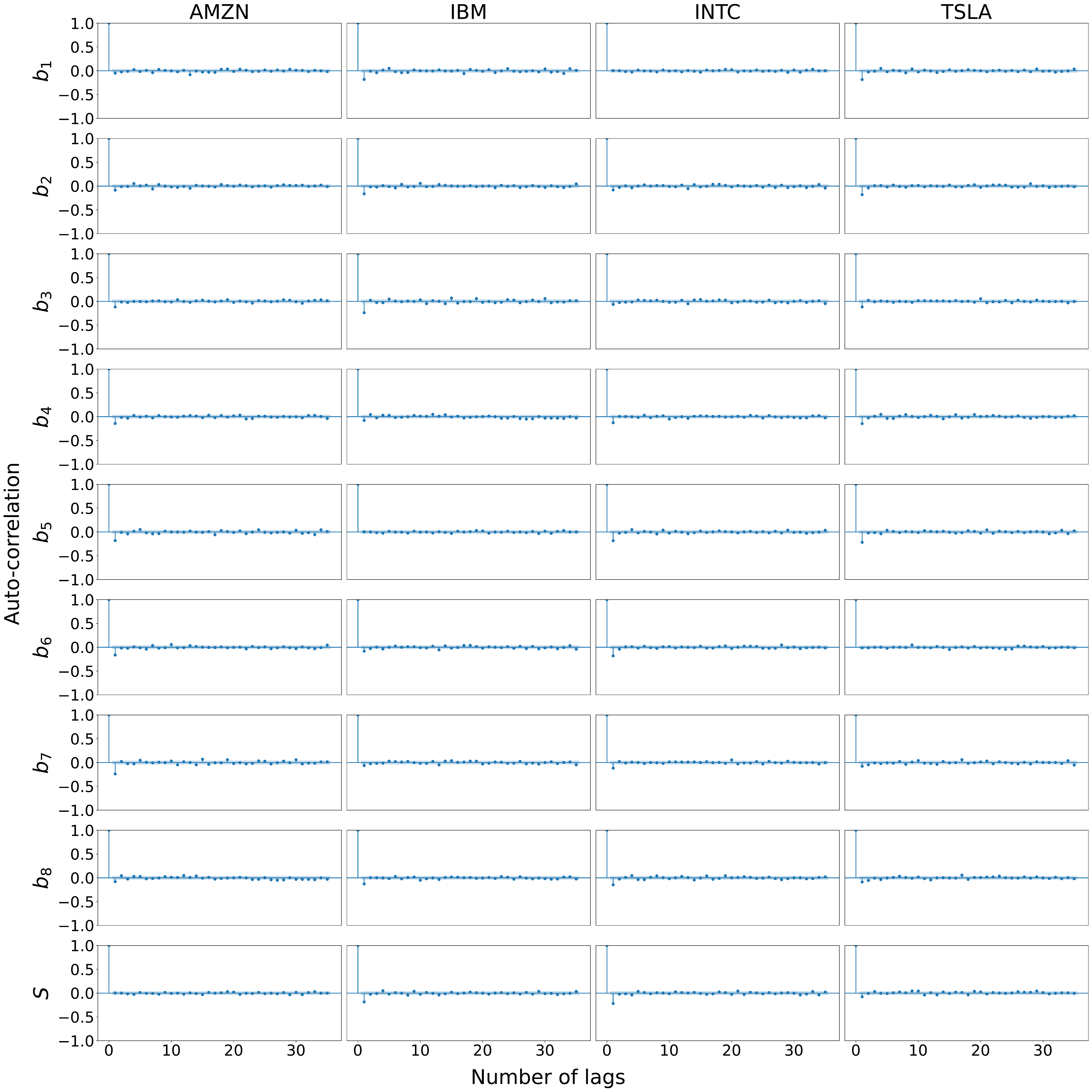}
    \caption{Auto-correlation of the transformed FPCCs and equity prices' PITs at the end of training when PIT penalty is part of the objective function.} 
    \label{fig:PIT_ACF}
\end{figure}

    \item Correlation: Comparing the correlation/cross-correlations between the observed and learnt time series is misleading for similar reasons above. The cross-correlation structure between the FPCCs and equity prices of the same asset, and/or  across assets, changes over time. To deal with this issue, while still having some measure of model misspecification in the correlation structure, we consider the sequence of PITs of the sum of  pairwise features. Given that our time series has $36$ features, we consider all $630$ `features' obtained by the pair-wise sum of  features. The learnt (conditional) one-step distribution of these `features' is also Gaussian with variance incorporating the correlation between the two raw features involved. Considering the sequence of PITs of these augmented `features', we then check if they are uniformly distributed via the Kolmogorov-Smirnov test. It is important to highlight that the objective function consists of the PIT penalty only on the univariate PITs, and not on the PITs of the augmented `features'.
    
    When the PIT penalty is minimized, the null hypothesis is rejected $44$ and $195$ out of $630$ times at the $1\%$ and $5\%$ significance level respectively. Note that these rejections are not caused just due to the correlation structure not being captured well, but may also be caused due to the learnt mean and variance haveing slight variations from the true ones. On the other hand, if we do not include the PIT penalty in the minimization problem, the null hypothesis is rejected $284$ and $506$ times at the $1\%$ and $5\%$ significance level respectively. This again indicates that including the PIT penalty enables to better capture the correlation structure over time.

    \begin{figure} [!ht]
    \centering
    \includegraphics[width=0.95\columnwidth]{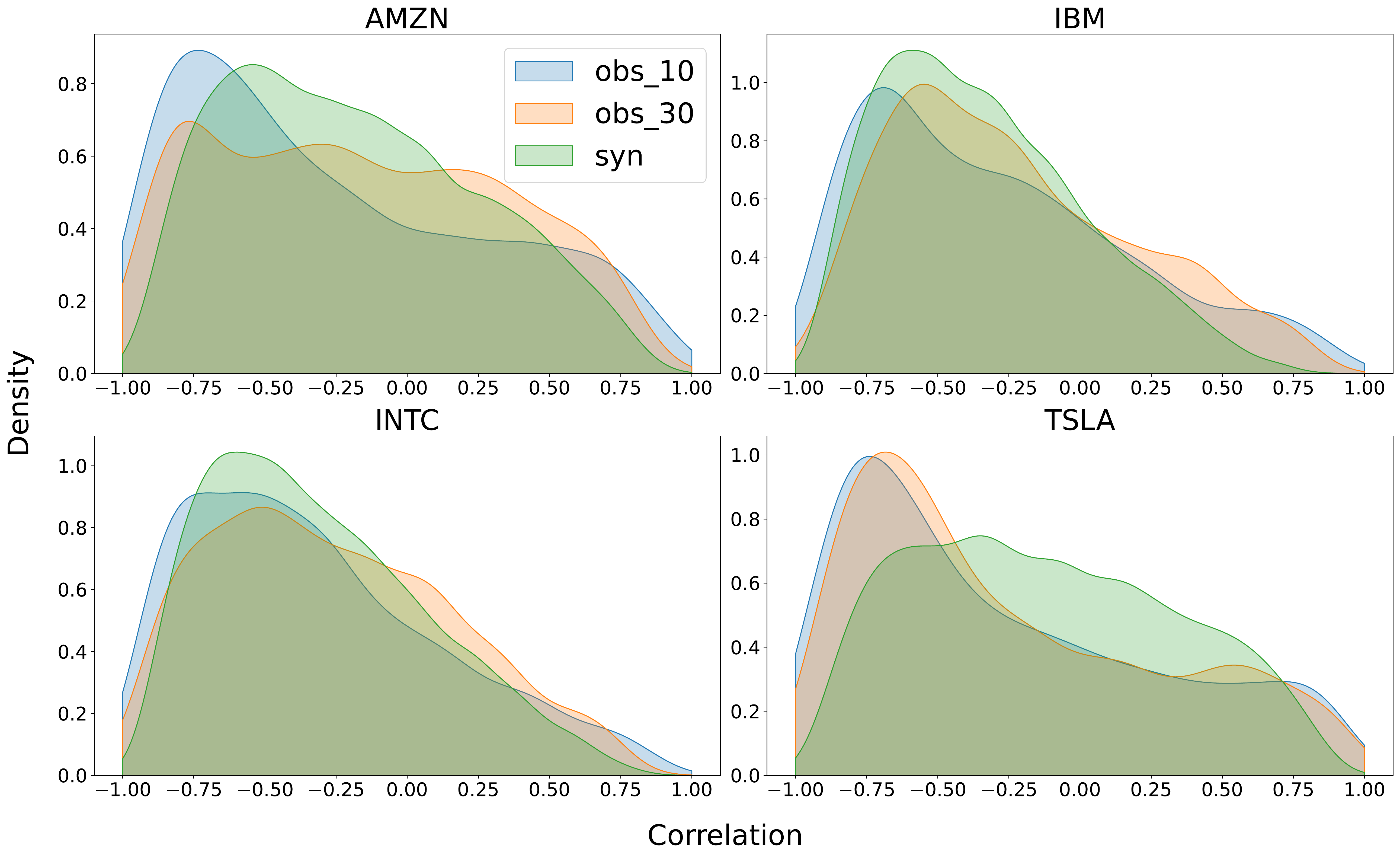}
    \caption{Distribution of the correlation between the first FPC and equity price of each asset for windows of length 10 (obs\_10) and 30 (obs\_30) in the observed data as well as that of 10,000 synthetic paths of length 30 (syn).} 
    \label{fig:fpc1_price_corr}
    \end{figure}
    
    Moreover, we  assess if the time-varying correlation structure is reflected in the synthetic scenarios as we would like them to be inclusive of all observed scenarios in the past. To this end, we consider the correlation between features over windows of length $10$ and $30$ in the training period in a rolling window fashion, sliding the window one day at a time. If we have $T$ observations, this would give us $T-9$ and $T-29$ values for the correlations between a pair. Given that ours is a conditional generative model, it is sensible to look at the correlations over small windows as these are the conditioning paths we consider. We now compare the kernel density estimate of these correlations with that of the correlations across the generated scenarios over a $30$ day period. Given that there are $630$ pairwise correlations, we present the results only for pairwise correlations between the first FPC and equity price for the same asset, and for the prices across assets. The results are presented in Figure \ref{fig:fpc1_price_corr} and \ref{fig:price_corr}. It is evidently clear that there is a larger density of negative correlations between the first FPC and the asset price in the observed data, something which is also reflected in the generated paths. Similarly, there is a higher positive correlation between the prices of distinct assets, once again captured by our generative model. In fact, the density plots for the correlations overlap significantly for both synthetic and observed data of different window lengths, and the model generates scenarios that are consistent with historical data -- both in terms of the values and the proportion of the correlation. This further provides evidence of our model learning the correlation structure really well.

    \begin{figure} [!ht]
    \centering
    \includegraphics[width=0.95\columnwidth]{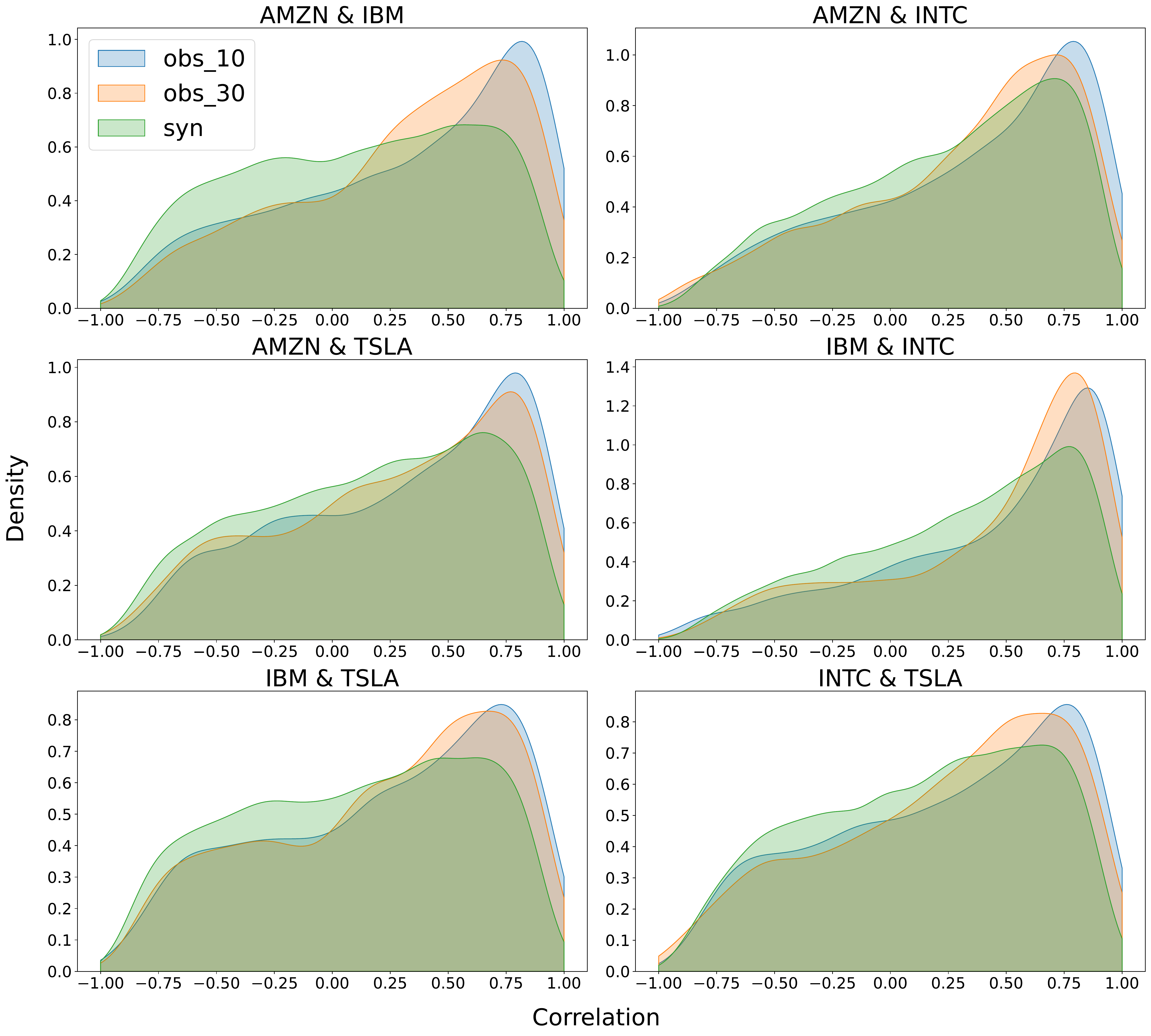}
    \caption{Distribution of the correlation between the equity prices of assets for windows of length 10 (obs\_10) and 30 (obs\_30) in the observed data as well as that of 10,000 synthetic paths of length 30 (syn).} 
    \label{fig:price_corr}
    \end{figure}    
\end{enumerate}

\end{document}